\documentclass[fleqn,usenatbib]{mnras}

\usepackage{newtxtext,newtxmath}

\usepackage[T1]{fontenc}

\DeclareRobustCommand{\VAN}[3]{#2}
\let\VANthebibliography\thebibliography
\def\thebibliography{\DeclareRobustCommand{\VAN}[3]{##3}\VANthebibliography}

\usepackage{graphicx}	
\usepackage{amsmath}	
\usepackage{amssymb}	
\usepackage{subcaption}
\usepackage{xcolor}
\definecolor{ao}{rgb}{0.0, 0.5, 0.0}
\definecolor{bv}{rgb}{0.54, 0.17, 0.89}
\definecolor{r}{rgb}{0.8, 0.0, 0.0}

\def\lya{\mbox{Ly\,{\sc $\alpha$}}}
\def\hi{\mbox{H\,{\sc i}}}

\def\Lya{Ly$\alpha\ $}
\def\wlya{$W_{\lambda}({\rm{Ly}\alpha})$}

\title[The LyC escape fraction at $z=3.5$]{The VANDELS survey: a measurement of the average Lyman-continuum escape fraction of star-forming galaxies at $\bmath z=3.5$}
\author[R. Begley et al.]{R. Begley$^{1}$\thanks{E-mail:rbeg@roe.ac.uk}, F. Cullen$^{1}$, R. J. McLure$^{1}$, J. S. Dunlop$^{1}$, A. Hall$^{1}$, A. C. Carnall$^{1}$, M. L. Hamadouche$^{1}$ \and D. J. McLeod$^{1}$, R. Amorín$^{2,3}$, A. Calabrò$^{4}$, A. Fontana$^{4}$, J. P. U. Fynbo$^{5}$, L. Guaita$^{4,6}$, N. P. Hathi$^{7}$, \and  P. Hibon$^{8}$, Z. Ji$^{9}$, M. Llerena$^{3}$, L. Pentericci$^{4}$, A. Saldana-Lopez$^{10}$, D. Schaerer$^{10}$, M. Talia$^{11,12}$, \and E. Vanzella$^{13}$, G. Zamorani$^{13}$\\\\
$^{1}$Institute for Astronomy, University of Edinburgh, Royal Observatory, Edinburgh EH9 3HJ\\
$^{2}$Instituto de Investigación Multidisciplinar en Ciencia y Tecnología, Universidad de La Serena, Raúl Bitrán, 1305 La Serena, Chile\\
$^{3}$Departamento de Física y Astronomía, Universidad de La Serena, Av. Juan Cisternas 1200 Norte, La Serena, Chile\\
$^{4}$INAF – Osservatorio Astronomico di Roma, via Frascati 33, 00078, Monteporzio Catone, Italy\\
$^{5}$The Cosmic Dawn Center, Niels Bohr Institute, University of Copenhagen, Juliane Maries Vej 30, DK-2100 Copen- hagen Ø, Denmark\\
$^{6}$Núcleo de Astronomía, Facultad de Ingeniería, Universidad Diego Portales, Av. Ejército 441, Santiago, Chile\\
$^{7}$Space Telescope Science Institute, 3700 San Martin Drive, Baltimore, MD 21218, USA\\
$^{8}$European Southern Observatory (ESO), Vitacura, Chile\\
$^{9}$Department of Astronomy, University of Massachusetts Amherst, 710 N. Pleasant St., Amherst, MA 01003, USA\\
$^{10}$Department of Astronomy, University of Geneva, 51 Chemin Pegasi, 1290 Versoix, Switzerland\\
$^{11}$INAF – Osservatorio Astronomico di Bologna, Via P. Gobetti 93/3, 40129 Bologna, Italy\\
$^{12}$University of Bologna, Department of Physics and Astronomy (DIFA) Via Gobetti 93/2- 40129, Bologna, Italy\\
$^{13}$INAF - OAS, Osservatorio di Astrofisica e Scienza dello Spazio di Bologna, via Gobetti 93/3, I-40129 Bologna, Italy \\
}

\date{Accepted 2022 April 11. Received 2022 April 11; in original form 2022 February 8}

\pubyear{2021}

\begin{document}
\label{firstpage}
\pagerange{\pageref{firstpage}--\pageref{lastpage}}
\maketitle

\begin{abstract}
 We present a study designed to measure the average Lyman-continuum escape fraction ($\langle f_{\rm esc}\rangle$) of star-forming galaxies at $z\simeq 3.5$. We assemble a sample of 148 galaxies from the VANDELS spectroscopic survey at $3.35~\leq~z_{\rm spec}~\leq~3.95$, selected to minimize line-of-sight contamination of their photometry. For this sample, we use ultra-deep, ground-based, $U-$band imaging and Hubble Space Telescope $V-$band imaging to robustly measure the distribution of $\mathcal{R_{\rm obs}}$ $=(L_{\rm LyC}/L_{\rm UV})_{\rm obs}$. We then model the $\mathcal{R_{\rm obs}}$  distribution as a function of $\langle f_{\rm esc}\rangle$, carefully accounting for attenuation by dust, the intergalactic medium and the circumgalactic medium. A maximum likelihood fit to the $\mathcal{R_{\rm obs}}$ distribution returns a best-fitting value of $\langle f_{\rm esc}\rangle =0.07^{+0.02}_{-0.02}$, a result confirmed using an alternative Bayesian inference technique (both techniques exclude $\langle f_{\rm esc}\rangle=0.0$ at $> 3\sigma$). By splitting our sample in two, we find evidence that $\langle f_{\rm esc}\rangle$ is positively correlated with Ly$\alpha$ equivalent width (\wlya), with high and low \wlya \ sub-samples returning values of $\langle f_{\rm esc}\rangle=0.12^{+0.06}_{-0.04}$ and $\langle f_{\rm esc} \rangle=0.02^{+0.02}_{-0.01}$, respectively. In contrast, we find evidence that $\langle f_{\rm esc}\rangle$ is anti-correlated with intrinsic UV luminosity and UV dust attenuation; with low UV luminosity and dust attenuation sub-samples both returning best fits in the range $0.10 \leq \langle f_{\rm esc}\rangle \leq 0.22$. We do not find a clear correlation between $f_{\rm esc}$ and galaxy stellar mass, suggesting stellar mass is not a primary indicator of $f_{\rm esc}$. Although larger samples are needed to further explore these trends, our results suggest that it is entirely plausible that the low dust, low-metallicity galaxies found at $z\geq 6$ will display the $\langle f_{\rm esc}\rangle\geq 0.1$ required to drive reionization.
\end{abstract}




\section{Introduction}

During the "Epoch of Reionization" (EoR), the Universe underwent a phase change in which the hydrogen gas in the intergalactic medium (IGM) was transformed from its early cold neutral state into the largely ionised IGM we see around us today. Current data indicate that the EoR spanned the approximate redshift range 
from $z \simeq 10-15$ down to $z \simeq 5 - 6$ (\citealt{robertson+15, robertson+21, bosman+21, goto+21}). However, the detailed progress and physical drivers of reionization currently remain highly uncertain and somewhat controversial, with some evidence supporting a late, short-lived, rapid reionization process \citep[e.g.][]{Mason18}, while other indicators favour a more gradual evolution of the IGM, commencing at much higher redshift \citep[e.g][]{Wu21}. 

Historically, the two main candidates for producing the bulk of the LyC photon budget required to achieve hydrogen reionization have been active galactic nuclei (AGN) and/or star-forming galaxies. However, with the number density of quasars and lower-luminosity AGN now known to fall rapidly at high redshift (\citealt{Aird15,Parsa18,McGreer18,Kulkarni19,Faisst21}), and recent constraints limiting the escape fraction of LyC photons in AGN to f$_{\rm esc}\ll$1 \citep{Iwata22}, early star-forming galaxies are now thought to be the primary source of ionising photons \citep{Chary16}. 

With ever improving measurements of the galaxy luminosity function at high redshift \citep{Bowler20,Harikane21}, it is becoming possible to track the progress of galaxy-driven reionization. However, doing so accurately requires reliable estimates of the production rate of LyC photons from early galaxies \citep[e.g.][]{tang+19}, and the average escape fraction ($\langle f_{\rm esc}\rangle$) of such photons into the IGM \citep[e.g.][]{Ocvirk21}.

A number of studies have attempted to use measurements of the evolving UV luminosity density produced by the early star-forming galaxy population to estimate what the $\langle f_{\rm esc}\rangle$ from young galaxies must be in order to deliver hydrogen reionization within the required time-frame \citep{bouwens+15,Bouwens21,finkelstein+15,finkelstein+19,robertson+13,robertson+15}. For example,  \cite{robertson+13} suggested that 
$\langle f_{\rm esc}\rangle$ in the range $10-20$ per cent is required, whereas the modelling of \cite{finkelstein+19} concluded that $\langle f_{\rm esc}\rangle \simeq 5$ per cent may suffice (helped by the gradual reduction in the optical depth $\tau$ to electron scattering derived from successive releases of data from Planck: now $\tau = 0.056 \pm 0.007$; \citealt{planck20}).

The inferred values for average $\langle f_{\rm esc}\rangle$ quoted by such studies are inevitably dependent on a number of empirical results and model assumptions, such as the assumed ionising photon production efficiency of early galaxies \citep[e.g.][]{eldridge+17}. One common additional assumption is that the production of LyC photons is dominated by the more numerous faint (and presumably metal poor) galaxies at such early times, however alternative assumptions can be explored. 

An example is presented in \cite{naidu+20}, who propose a model allowing $f_{\rm esc}$ to vary based on star-formation rate surface density. In contrast to requiring a low-to-moderate $\langle f_{\rm esc}\rangle$ across the entire galaxy population, their work suggests that $\gtrsim$80 per cent of the ionising photon budget may be accounted for by rarer, more massive galaxies with higher than average $f_{\rm esc}$. Such uncertainties over the production and escape of LyC photons from EoR galaxies arise because direct measurements of the LyC emission from these objects are impossible. Therefore, with the aim of studying galaxies that are most directly analogous to those that drove reionization, many studies have focused on searching for LyC leakers at $3 \leq z \leq4$, where the level of IGM transmission still allows the direct detection of LyC emission \citep{inoue+14}.

Some of the earliest successes in such searches have come from deep rest-frame UV spectroscopy, both through targeted surveys such as KLCS \citep{steidel+18} and from serendipitous discoveries. The former has resulted in 13 secure detections of LyC emission \citep{pahl+21}, including the LyC leaker Q1549-C25 first reported in \cite{shapley+16}. Other discoveries, both serendipitous and targeted, include such notable objects as \textit{Ion 1-3} (\citealt{vanzella+12a, vanzella+16a, deBarros+16,vanzella+18,ji+20}) and the Sunburst galaxy (\citealt{rivera+19,vanzella+21}). However, with less than 20 spectroscopically confirmed LyC leakers discovered to date at intermediate redshifts, the available sample of such sources remains small.

Constraints on the average escape fraction can also be derived from detailed analyses of larger samples that do not feature significant individual LyC detections, for example $\langle f_{\rm esc}\rangle=0.06\pm0.01$ \citep{pahl+21}. However, even with relatively large samples of galaxies, it is still difficult to achieve robust constraints on the typical level of LyC flux at intermediate redshifts, as shown in the meta-analysis by \cite{mestric+21}. Their work collates literature results from the last $\sim$ 20 years, finding that many studies were only able to  derive upper limits on $\langle f_{\rm esc}\rangle$, even from deep spectroscopic observations.

As a potentially efficient alternative to spectroscopic searches, a growing number of studies have sought to use narrow and/or broadband imaging to hunt for LyC emitting galaxies. The main observational requirement for such searches is the availability of deep \textit{U-}band imaging, which a number of studies have obtained via programmes such as CLAUDS \citep{mestric+20}, and LACES \citep{fletcher+19}, resulting in a number of likely LyC emitting candidates. As with spectroscopic studies, high angular-resolution imaging (effectively from {\it HST}) is required to robustly decontaminate samples of potential LyC leakers \citep{siana+15}. 
Regardless of their ability to unveil new candidate LyC emitting galaxies, these deep \textit{U-}band imaging surveys have generally only been able to place upper limits on $\langle f_{\rm esc}\rangle$ across their full galaxy samples (e.g. \citealt{guiata+16, grazian+17, saxena+21}). 
	
In the past decade a significant amount of effort has also been directed towards exploring which galaxy properties are correlated with, and therefore can be utilised as indirect indicators of, the level of leaking ionising radiation.
To date, the most promising such indicators for $f_{\rm esc}$ are tied to Ly$\alpha$ emission. Most recently, \cite{pahl+21} confirmed the positive correlation between increased $f_{\rm esc}$ and Ly$\alpha$ equivalent width ($\rm{W_{\lambda}(Ly\alpha)}$), previously found by \cite{steidel+18}. This statistical link is physically supported by simulations, which show that both ionising continuum flux and Ly$\alpha$ line emission can escape through the same ionised channels in the interstellar medium (ISM) (e.g. \citealt{kimm+14, wise+14}). 
	
However, the case for $\rm{W_{\lambda}(Ly\alpha)}$ as a clean proxy for LyC leakage is not clear cut (e.g. \citealt{mostardi+13}).
With a host of properties able to alter the transmission of both Ly$\alpha$ and ionising continuum photons, such as geometry and gas kinematics \citep{dijkstra+16}, any relationship between $\rm{W_{\lambda}(Ly\alpha)}$ and $f_{\rm esc}$ is undoubtedly complex. Conflicting results also exist for the connection between other galaxy properties and $f_{\rm esc}$, such as galaxy stellar mass and UV magnitude (e.g. \citealt{fletcher+19, izotov+21, pahl+21}), both of which are particularly important in the ongoing debate over which galaxies provided the bulk of the ionising photon budget in the EoR.

In addition to Ly$\alpha$, a number of other rest-frame UV/optical spectral features, accessible by {\it JWST} for galaxies within the EoR, have also been scrutinised as potential indicators of LyC leakage (\citealt{nakajima+14,ramambason+20,katz+20,mauerhofer+21}). 
In particular, the usefulness of [\mbox{O\,{\sc iii}}] line emission from galaxies ($\rm{W_{\lambda}( [\mbox{O\,{\sc iii}}])}$ and the [\mbox{O\,{\sc iii}}]/ [\mbox{O\,{\sc ii}}] ratio, hereafter O32), has been much explored, due to their association with recent bursts of star-formation activity and increased ionising photon production efficiency (\citealt{vanzella+16b, izotov+18, tang+19, tang+21a, endsley+21, tang+21b}). Indeed, more extreme  [\mbox{O\,{\sc iii}}] properties are usually attributed to the presence of density-bounded \mbox{H\,{\sc ii}} regions \citep{kewley+19}, from which LyC photons are thought to escape \citep{jaskot+19}. However, as with other proposed proxy indicators of LyC leakage, the link between  $\rm{W_{\lambda}( [\mbox{O\,{\sc iii}}])}$, O32, and $f_{\rm esc}$ is not conclusive (e.g. \citealt{naidu+20, saxena+21}).

The analysis by \cite{nakajima+20} suggests that high O32 is a requirement for high $f_{\rm esc}$, but that not all galaxies with high O32 are necessarily LyC leakers (a situation that is mirrored by the relationship between O32 and  Ly$\alpha$  emission; \citealt{tang+21a}). That no single measure has proven to be a clear and universal indicator of LyC leakage highlights the complexity of the underlying physics, in which anisotropic or time-evolving leakage may play a significant role, potentially explaining apparently inconsistent results (\citealt{cen+15,steidel+18,fletcher+19}). The lack of a clear consensus further bolsters the case for studying larger sample sizes with varied and complete datasets and/or using alternative methodologies \citep[see also][for constraints derived using gamma-ray bursts and galaxy-IGM cross-correlations, respectively]{tanvir2019,meyer2020}.

To try to advance this situation, in this study we have assembled a large sample of 
star-forming galaxies at $3.35 \leq z_{\rm spec} \leq 3.95$ from the ultra-deep VANDELS spectroscopic survey (\citealt{mclure+18,pentericci+18,garilli+21}). We utilise deep VLT/VIMOS $U$-band imaging to probe LyC emission ($\lambda_{{\rm rest}}\simeq820$ \AA), along with high resolution {\it HST} imaging to measure non-ionising UV fluxes ($\lambda_{{\rm rest}}\simeq1300$ \AA) and effectively clean the sample from line-of-sight contamination. We have calibrated the imaging with additional astrometric corrections, and have undertaken sophisticated depth determinations to ensure that our derived photometric uncertainties are robust. 
We compare the observed distribution of ionising to non-ionising flux ratios with simulated ratios from a realistic model which is based on physically motivated and/or empirically measured inputs, and includes an accurate treatment of both IGM and circumgalactic medium (CGM) transmission via Monte Carlo sightline simulations. From this thorough analysis, for the first time via a broadband imaging-based approach, we provide a statistical measurement ($\geq3\sigma$) of the sample-averaged absolute escape fraction at $z \simeq 3.5$. 

The paper is structured as follows. In Section \ref{sec:data} we describe the datasets used in this study, focusing on sample selection and cleaning, together with the additional calibration steps we have employed to extract robust photometry and accurately measure the LyC to non-ionising UV flux ratios. In Section \ref{sec:model}, we describe the construction of a model that can relate $\langle f_{\rm esc}\rangle$ to the observed LyC to non-ionising UV flux ratios, including the careful treatment of attenuation by dust, the IGM and CGM. Our constraints on  $\langle f_{\rm esc} \rangle$ for the full sample are presented in Section \ref{sec:results}, where we also explore potential correlations between $\langle f_{\rm esc}\rangle$ and Ly$\alpha$ equivalent width, UV luminosity, stellar mass and UV dust attenuation. We discuss the significance of our results in Section \ref{sec:discussion}, before summarising our conclusions in Section \ref{sec:conclusions}. Throughout the paper we adopt cosmological parameters $H_{0}$~=~$\mathrm{70 \ {km} \ {s^{-1}} \ {Mpc^{-1}}}$, $\mathrm{\Omega_{m}}$ = 0.3 and $\mathrm{\Omega_{\Lambda}}$ = 0.7. All magnitudes are quoted in the AB system (Oke \& Gunn 1983), and unless otherwise stated, we refer to the absolute escape fraction as simply the escape fraction.

\section{Data and sample selection}\label{sec:data}
The constraints on $\langle f_{\rm esc}\rangle$ derived
in this work fundamentally rely on accurately measuring the observed ratio of LyC to non-ionizing UV flux in a sample of star-forming galaxies at $z\simeq 3.5$. 

The three key datasets necessary to perform this experiment are all publicly available.
A suitable sample of spectroscopically confirmed star-forming galaxies has recently been provided in the CDFS by the final data release (DR4) of the VANDELS ESO public spectroscopic
survey \citep{garilli+21}. Moreover, the necessary measurements of the observed LyC flux are provided by the publicly available, ultra-deep, $U-$band imaging of the CDFS
presented by \cite{nonino+09}. Finally, the necessary measurements of the non-ionizing UV flux are provided by the {\it HST} ACS F606W, hereafter $V_{606}$, imaging of the CDFS,
released as part of version 2.0 of the Hubble Legacy Field programme\footnote{\tt https://archive.stsci.edu/prepds/hlf/}(\citealt{whitaker+19}).
In this section we fully describe the sample selection process, together with the steps taken to extact robust photometry from
the ground-based and {\it HST} imaging.

\subsection{The VANDELS survey}
The sample of star-forming galaxies utilized in this work is drawn exclusively from the final data release (DR4) of the VANDELS ESO public spectroscopy
survey (\citealt{mclure+18}; \citealt{pentericci+18}; \citealt{garilli+21}). The VANDELS survey obtained ultra-deep (20-80 hours of integration),
red optical ($4800<\lambda_{\rm obs}<10200$ \AA) spectra for a sample of 2087 galaxies with the VIMOS spectrograph on the VLT. The primary VANDELS sample,
accounting for $83$ per cent of the spectroscopic targets, consisted of galaxies on the star-forming main sequence \citep[e.g.][]{daddi+07}
within the redshift interval $2.4 \leq z_{\rm phot} \leq 6.4$. Full details of the survey design can be found in \cite{mclure+18} and a detailed description of the data
reduction, data quality assurance and spectroscopic redshift determinations can be found in \cite{pentericci+18} and \cite{garilli+21}.

The initial sample selected for this work consists of 242 VANDELS DR4 star-forming galaxies within the CDFS\footnote{A similar number of suitable VANDELS DR4 galaxies are available in the UDS survey field, however the UDS currently lacks the necessary ultra-deep $U-$band imaging data.}, with high-quality spectroscopic redshifts ($z_{\rm flag}=3$ or 4)
within the interval $3.35 \leq z_{\rm spec} \leq 3.95$. We note here that the spectroscopic redshifts for VANDELS galaxies with quality flags $z_{\rm flag}=3$ or 4 are derived from 
multiple spectral features and the analysis presented by \cite{garilli+21} confirms that they are reliable at the $99$ per cent level.
Each galaxy had an associated stellar mass derived from multi-wavelength broad band photometry using the SED-fitting code \textsc{BAGPIPES} \citep{,carnall+18,carnall+19} as described in \citet{garilli+21}, and a measured \lya{} equivalent width (\wlya) following the method outlined in \citet{cullen+20} \citep[following][]{kornei+10}. The UV magnitude ($M_{\rm UV}$) of each galaxy is calculated based on the VANDELS spectra and available photometry, following the method outlined in Section \ref{subsec:auv}.

The low-redshift limit at $z_{\rm spec}= 3.35$ was imposed to ensure that the $U-$band filter used for the VIMOS imaging only samples rest-frame wavelengths
short-ward of the Lyman limit. In contrast, the high-redshift limit was determined based on a simulation of the combined impact of the IGM and CGM on the transmission of LyC photons. This indicated that $z_{\rm spec}=3.95$ was the redshift at which
the increasing opacity of the IGM+CGM outweighed the improved signal-to-noise provided by a larger sample size.
We note that \cite{vanzella+10a} reached the same conclusion regarding the optimal redshift window for detecting potential LyC emission
in an earlier study using the same $U-$band imaging data.

    \begin{figure}
        \centerline{\includegraphics[width=\columnwidth]{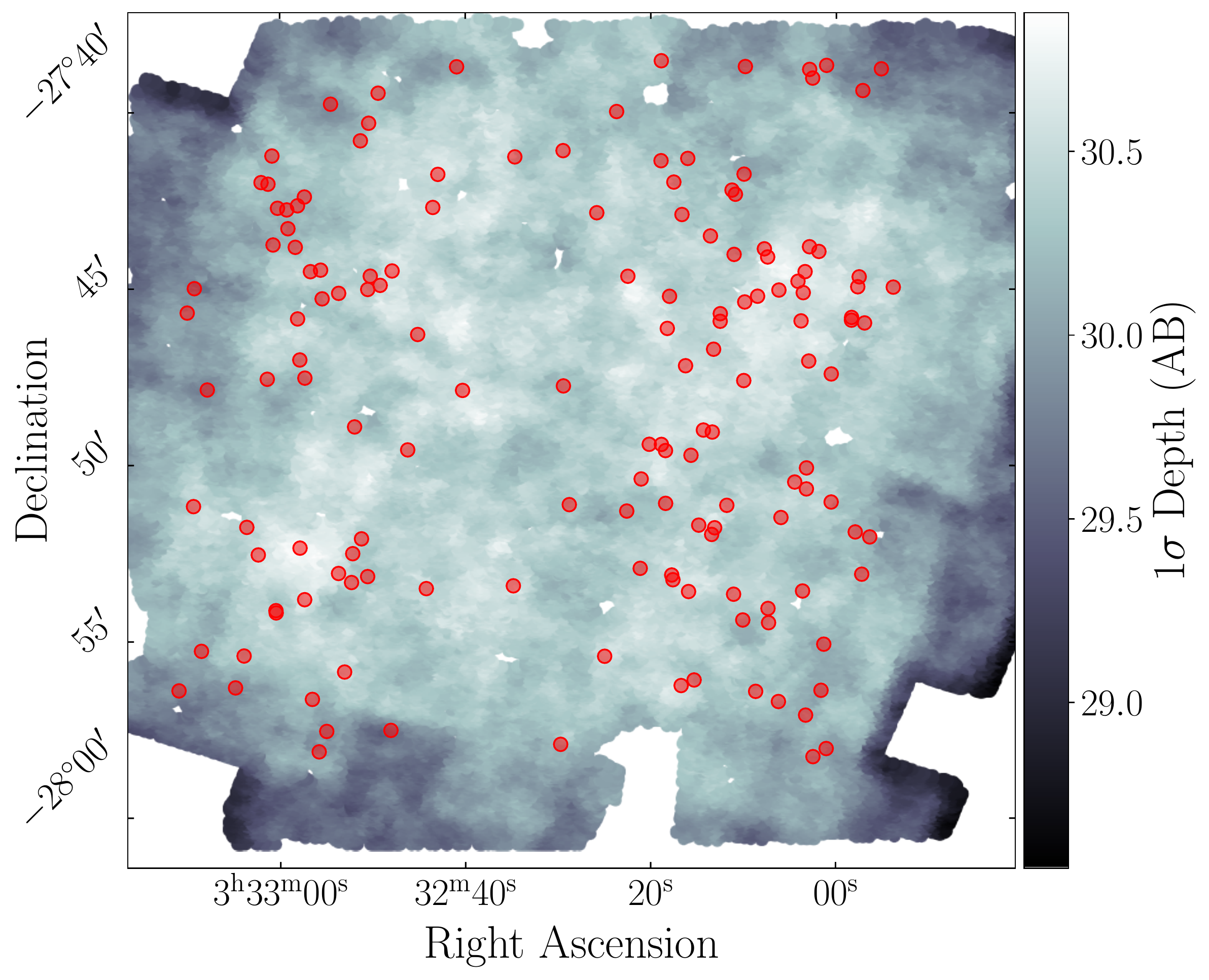}}
        \caption{Depth map for the CDFS $U-$band mosaic, as measured within photometric apertures with a diameter of 1.2 arcsec. The locations of the 148 galaxies within our final star-forming galaxy sample are shown as filled red circles. Mapping the spatially varying depth allows robust photometric uncertainties to be allocated to each galaxy, based on its position.}
        \label{fig:UdepthMap}
    \end{figure}

\subsection{Imaging data}
This study makes use of the $U-$band imaging of the CDFS field obtained with VLT+VIMOS by \cite{nonino+09} and the coincident $V_{606}$ imaging obtained with \emph{HST}. Fitting with the {\sc psfex} software
package \citep{Bertin13} demonstrated that the PSF of the $U-$band mosaic shows little spatial variation. Over the area of the mosaic where the VANDELS star-forming galaxies are located, we find a median FWHM of 0.79 arcsec and a maximum FWHM of 0.80 arcsec. As a result, throughout this paper we measure photometry within circular apertures with a diameter of 1.2 arcsec, in order to maximize the signal-to-noise ratio for compact sources \citep[e.g.][]{Brammer16}. 

\subsubsection{PSF homogenisation}
Our adopted method for determining $\langle f_{\rm esc} \rangle$ relies on an accurate measurement of the LyC to non-ionising UV flux ratio; effectively the $U-V_{606}$ colour. This measurement clearly relies on the $U-$band and $V_{606}-$band aperture photometry capturing the same fraction of total flux for each object. To meet this requirement, the $V_{606}$ image was PSF-homogenised to the $U-$band image using a convolution kernel generated by {\sc Photutils}, based on stacks of isolated stars. Following PSF homogenisation, a curve of growth analysis confirmed that the enclosed relative flux within an 1.2-arcsec diameter aperture on the $U-$band and $V_{606}$-band images matched to within $\pm 2$ per cent.

\subsubsection{Astrometry calibration}
In addition to PSF homogenisation, the measurement of an accurate $U-V_{606}$ colour requires any astrometric shifts between the $U-$band and $V_{606}-$band images to be minimised.
To address this issue we selected a catalogue of bright sources, detected in both images with S/N $\geq 8\sigma$, with positions that matched within a tolerance of 0.5 arcsec.
This catalogue revealed that the median astrometry offset between the two images was $\Delta \alpha=0.133$ arcsec.

By applying a spatially varying correction to the $U-$band astrometry, based on the median off-sets of the nearest 200 bright objects, it was possible to reduce the
median astrometry offset to $\Delta \alpha=0.08$ arcsec.
This improvement in astrometric accuracy allowed us to extract robust $U-$band photometry at the measured $V_{606}$ centroids, without astrometric shifts contributing
significantly to the uncertainty in the $U-V_{606}$ colours.

\subsubsection{Sky subtraction and depth analysis}\label{section:photometry}
We adopted a two-step process to address the issues of sky subtraction and the positionally varying depth of the $U-$band imaging.
The first step was to subtract a low-order, two-dimensional, sky-background fit to the $U-$band image with {\sc photutils},
using a dilated segmentation map to exclude objects from the fit. Following this global sky-subtraction, a second step was employed to deal with any remaining local variations.
This step involved creating a dense grid of non-overlapping blank-sky apertures, each with a diameter of 1.2 arcsec.
For each galaxy in our final sample, the $U-$band photometry was measured within an aperture with a diameter of 1.2 arcsec, centred on the measured $V_{606}$ centroid,
with the median flux of the nearest 200 blank-sky apertures taken as the local sky-background estimate.
The corresponding value of $\sigma_{\rm MAD}$ measured from the flux distribution of the nearest 200 blank-sky apertures was adopted as the local $1\sigma$ depth estimate.
An identical procedure was followed to measure and quantify the $V_{606}-$band photometry extracted from the PSF-homogenised $V_{606}$ image.

A depth map, illustrating the spatial variation in sensitivity of the $U-$band image, is shown in Fig.~\ref{fig:UdepthMap}. Within the region occupied by our final galaxy sample,
we calculate a global median $1\sigma$ depth of $m_{1\sigma}=30.4$. Although there is clearly spatial variation in the depth of the $U-$band image,
$\geq 90$ per cent of our final galaxy sample lie in regions with $m_{1\sigma}\geq 30.1$. Having an accurate measurement of the spatially varying depth allows us to
allocate robust flux errors to each object as a function of their position.

\subsection{Final sample selection}

\begin{figure}
        \centerline{\includegraphics[width=0.925\columnwidth]{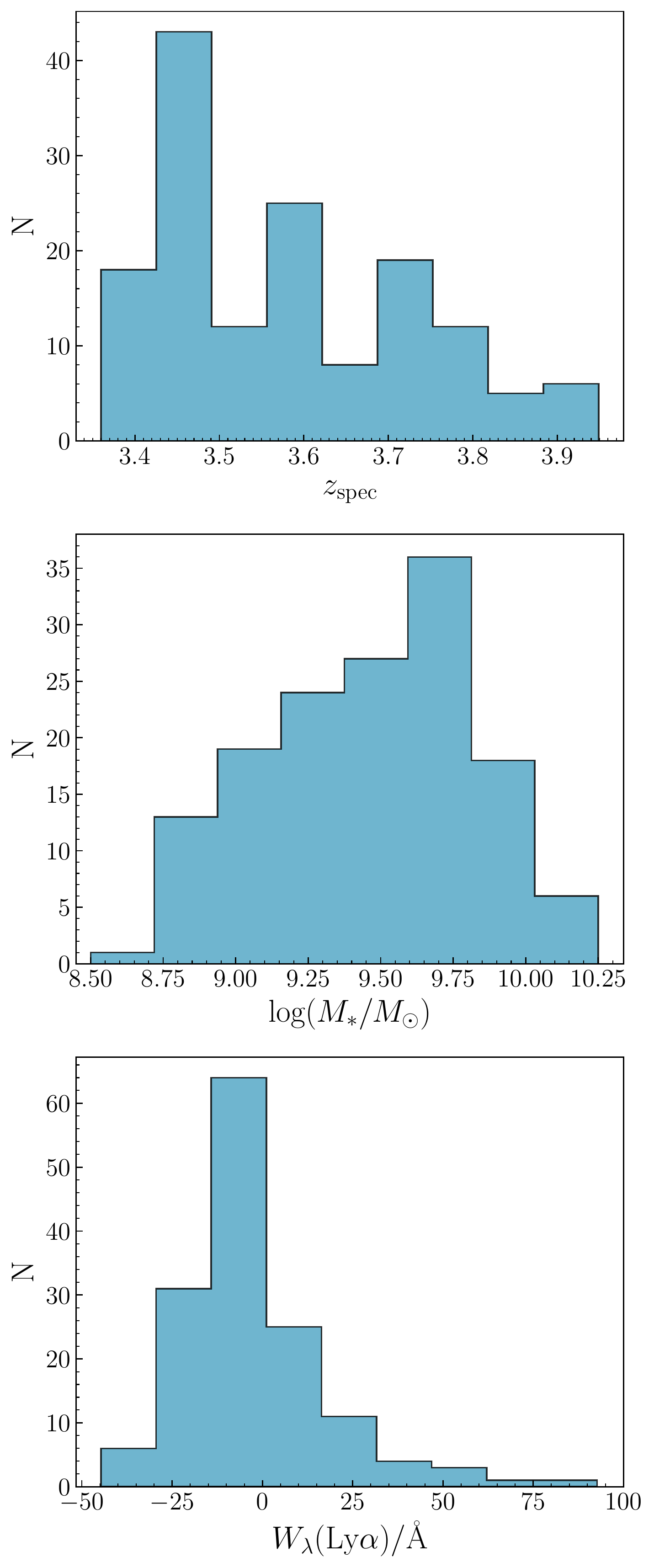}}
        \caption{The distribution of spectroscopic redshift (top), stellar mass (middle) and Ly$\alpha$ equivalent width (bottom) for our final sample 148 of star-forming galaxies. The median values of the three properties shown are $z_{\rm spec}=3.58$, $\log(M_{\star}/{\rm M}_{\odot})=9.5$ and W$_{\lambda}$(Ly$\alpha)=-6$ \AA\ (see § \ref{subsec:sample_splits}).}
        \label{fig:sampleProperties}
\end{figure}

Unfortunately, the whole initial sample of 242 star-forming galaxies is not suitable for constraining the escape fraction of LyC photons, primarily due to the potential for significant contamination
of the ground-based $U-$band photometry by flux from nearby companion objects.
It was therefore necessary to clean our initial sample for potential contaminants, as described below.

\subsubsection{Line-of-sight contamination in the imaging data}
 As discussed above, based on the $\simeq0.8$ arcsec FWHM of the $U-$band PSF, we adopt photometric apertures
with a diameter of 1.2 arcsec. Therefore, the first stage in cleaning the sample involved visually inspecting $U-$band cutouts of each object and removing all objects that displayed any level of contaminating flux from nearby companion objects within a radius of 0.6 arcsec.
From the initial sample, 23 objects were excluded due to having $U-$band photometric apertures that were unambiguously contaminated by flux from nearby objects, leaving a sample of 219 remaining objects.

The second stage of the cleaning process exploited the high-spatial-resolution $V_{606}$ imaging to identify small angular separation contaminants ($r<0.6$ arcsec)
that could not be identified from the low-spatial resolution $U-$band imaging. The third stage of the cleaning process made use of true-colour images constructed from the other
available {\it HST} ACS imaging data (i.e. F435W, F775W, F850LP)\footnote{ACS F850LP and F606W imaging was available for all galaxies, with additional F775W and F435W imaging available for 70 per cent of the sample.} in order to exclude those extended objects that visually displayed strong colour gradients, potentially indicative of
line-of-sight projections of objects at different redshifts. In total, based on the high-spatial resolution {\it HST} imaging, we excluded a further 34 objects,
leaving a sample of 185 objects.

For a further 32 galaxies, it was not possible to state unambiguously that they contain no contaminating flux within the photometric aperture through a combination of the three previous cleaning stages. As a result, these galaxies were classed as {\it potentially} contaminated, and given our conservative approach, we also excluded these objects, leaving a sample of 153.

\begin{figure}
        \centerline{\includegraphics[width=0.92\columnwidth]{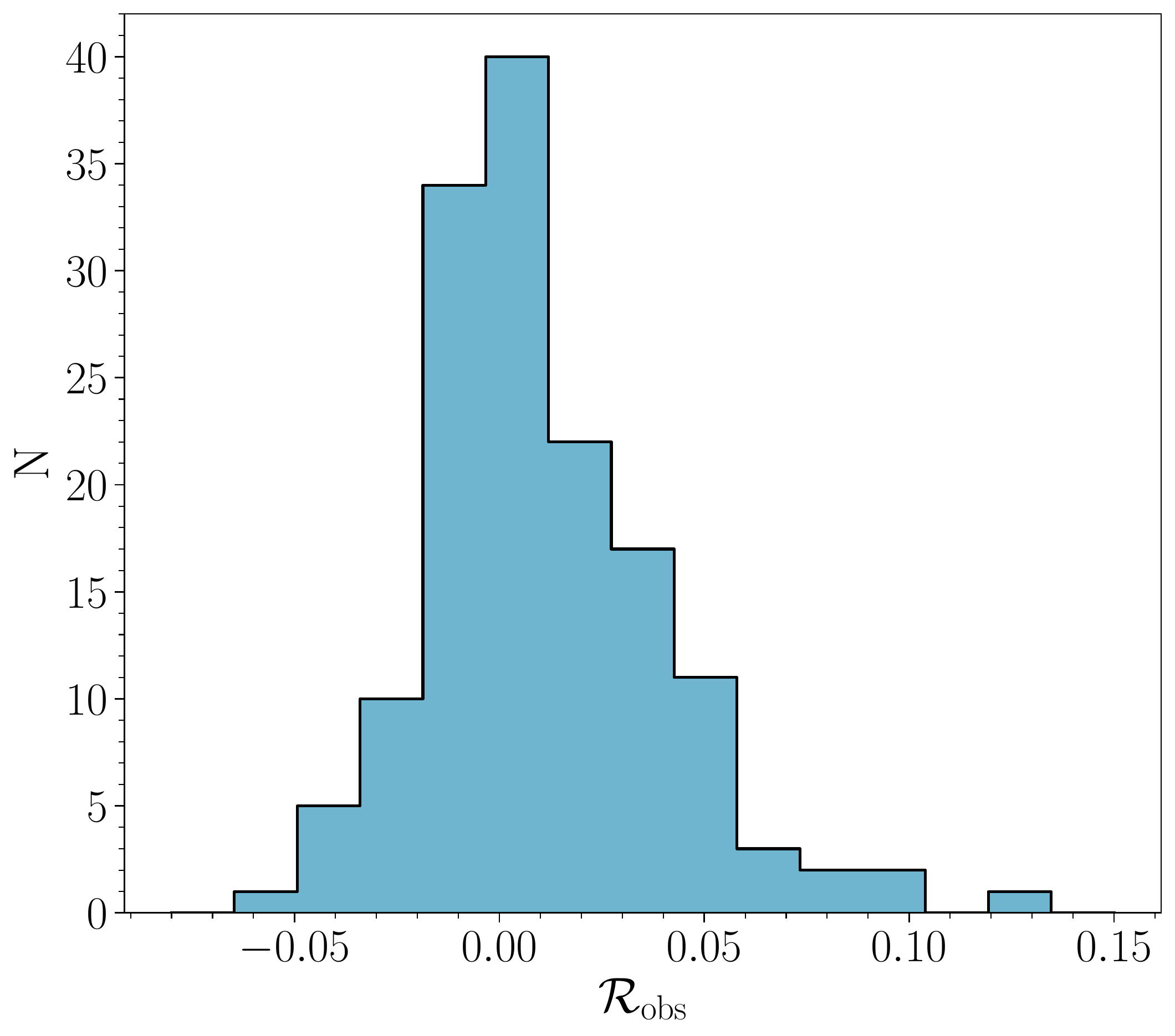}}
        \caption{The $\mathcal{R_{\rm obs}}$ distribution of our final sample of 148 star-forming galaxies, derived in Section \ref{sec:data}. It is the centroid and overall shape of this distribution that provides the fundamental observational constraint on $\langle f_{\rm esc} \rangle$.}
        \label{fig:observedratio}
\end{figure}

\subsubsection{AGN contamination}
The final stage of cleaning the sample involved excluding potential AGN. This process was based on the identification of sources within
the 7Ms Chandra X-ray catalogue of the E-CDFS \citep{Luo17}, which covers an area including the full VANDELS sample in the CDFS.
We excluded a further five objects as potential AGN, all of which could be associated with high SNR detections in the 7Ms X-ray catalogue, within an angular separation
of 1.1 arcsec.

\subsubsection{Final galaxy sample}
Following the exclusion of potential AGN, the final sample of galaxies 
consists of 148 star-forming galaxies within the
redshift interval $3.35 \leq z_{\rm spec} \leq 3.95$. Our conservative approach to cleaning excluded a total of 94 objects from the initial sample ($39$ per cent), primarily on the basis of potential photometric contamination from nearby objects.
The redshift, stellar mass and \wlya \ distributions of our final sample are shown in Fig. \ref{fig:sampleProperties}.
    
\subsubsection{The LyC to non-ionising UV flux ratio}
As discussed previously, the observational constraint on $f_{\rm esc}$ for a given galaxy is derived from the the LyC to non-ionising UV flux ratio:
\begin{equation}
  \mathcal{R_{\rm obs}}=\left(\frac{L_{\rm LyC}}{L_{\rm UV}}\right)_{\rm obs}=
  \left(\frac{\langle f_{\rm U}\rangle}{\langle f_{\rm V}\rangle}\right)_{\rm obs},
\end{equation}
where $\langle f_{\rm U}\rangle$ and $\langle f_{\rm V}\rangle$ are the flux densities per unit frequency
measured within 1.2-arcsec diameter apertures on the $U-$band and PSF-homogenised $V_{606}-$band images, respectively. In the next section, we describe the technique we have adopted to model the $\mathcal{R_{\rm obs}}$ distribution of the final sample (see Fig. \ref{fig:observedratio}), and thereby measure the value of $\langle f_{\rm esc}\rangle$. 
However, it is worth noting that it is $\mathcal{R_{\rm obs}}$ that is the fundamental observable and that, after correcting for the effects of the IGM and CGM, it is $\mathcal{R_{\rm obs}}$ that is directly related to a galaxy's total ionising emissivity. The subsequent conversion between $\mathcal{R_{\rm obs}}$ and $f_{\rm esc}$ is inevitably more model dependent, a fact that is worth remembering when comparing the values of $f_{\rm esc}$ derived from different studies.

\begin{figure*}
    \begin{subfigure}[b]{1\columnwidth}
        \includegraphics[width=\textwidth]{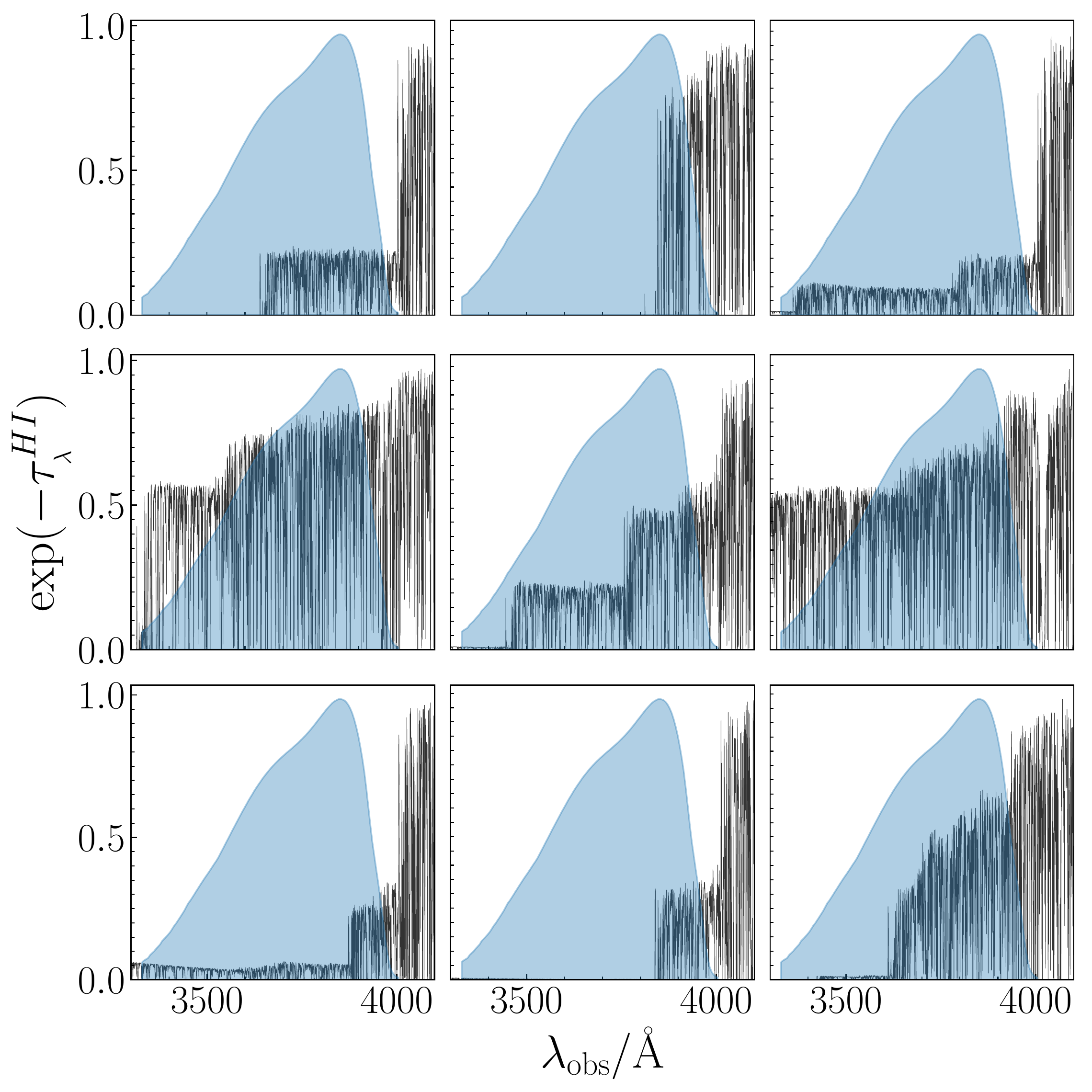}
    \end{subfigure}
    \hfill
    \begin{subfigure}[b]{1\columnwidth}
        \includegraphics[width=\textwidth]{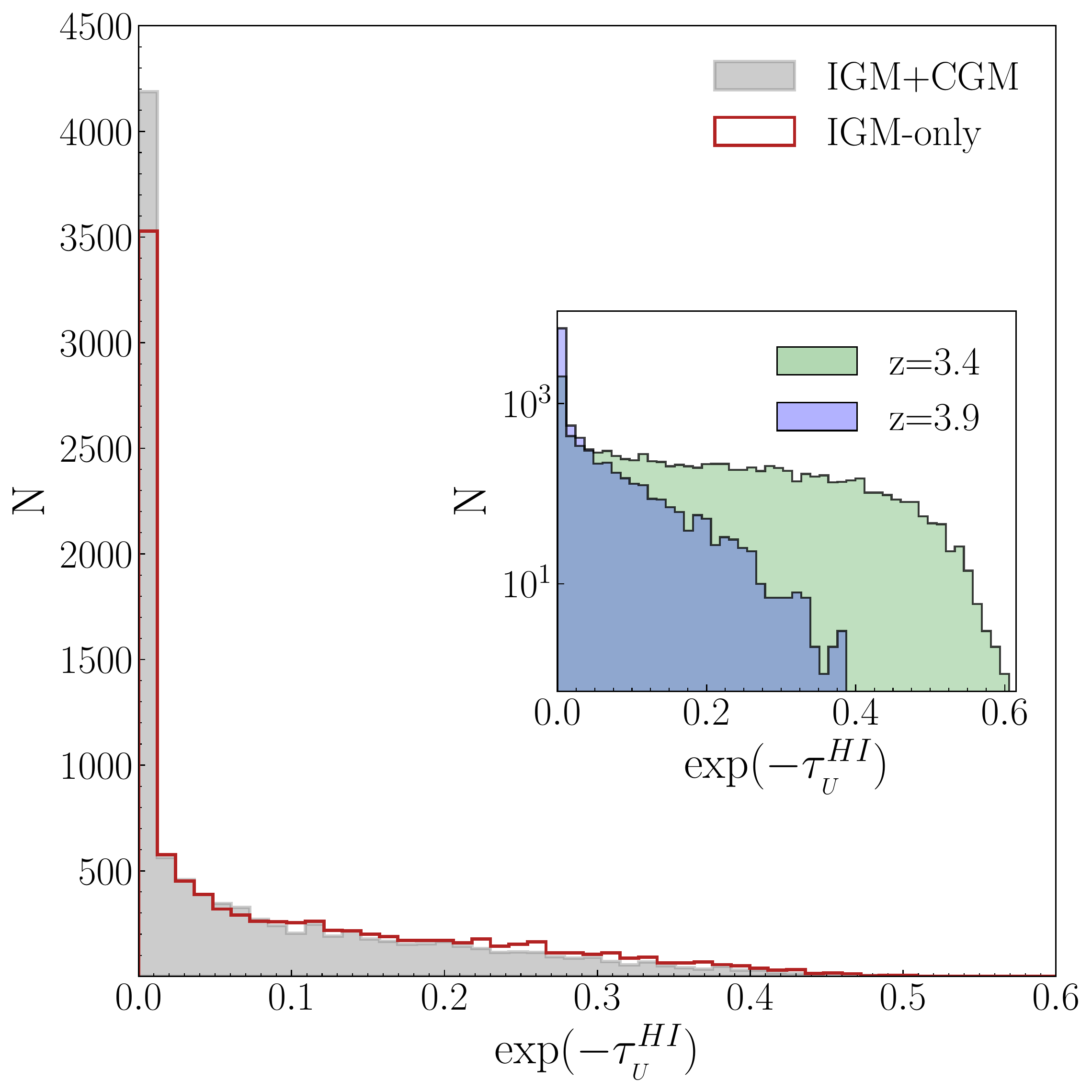}
    \end{subfigure}
    \caption{To accurately account for the fluctuating optical depth facing ionising photons from a combination of the IGM and CGM, we generated a large number of individual sight-line transmission curves as a function of redshift, as detailed in Section \ref{subsec:igmcgm_distributions}. \textit{Left panel:} Nine random IGM+CGM transmission curves generated at z=3.4 (black) over the observed wavelength range for LyC flux (the Lyman-limit lies at $\lambda_{\rm obs}=4013$\,\AA\,at $z=3.4$), illustrating the strong variation across different sight lines. The blue shading in the background of each panel shows the transmission of the $U-$band filter.
    \textit{Right panel:} Histograms showing the average transmission from 10,000 sight lines generated at $z=3.55$, calculated by integrating the transmission curve for each sight line through the \textit{U-}band filter. The filled grey histogram corresponds to sight lines accounting for \textit{both} the IGM and CGM, while the red histogram shows IGM-only sight lines. The importance of incorporating the CGM contribution is highlighted by the significant increase in the relative number of sight lines with $\langle{e^{-\tau}_{\rm U}}\rangle\simeq0$. The inset panel shows the $\langle{e^{-\tau}_{\rm U}}\rangle$ distribution of sight lines at z=3.4 (green) and z=3.9 (blue), highlighting the increasing optical depth at higher redshifts.}
    \label{fig:igm_cgm_transmission}
    
\end{figure*}

\section{Modelling the  $\mathcal{R_{\rm obs}}$ probability distribution }\label{sec:model}

Armed with the $\mathcal{R_{\rm obs}}$ distribution we can then proceeded to estimate $\langle f_{\rm esc}\rangle$ for the final sample of 148 star-forming galaxies. To do this we first constructed a generative model to predict the probability distribution of $\mathcal{R_{\rm obs}}$ for a given object, as a function of $f_{\rm esc}$.
The basic equation relating $\mathcal{R_{\rm obs}}$ to $f_{\rm esc}$ is:
\begin{equation}\label{eq:fesc_eq}
\mathcal{R_{\rm obs}} = f_{\rm esc} \times \mathcal{R_{\rm int}} \times e^{-\tau_{\lambda}^{\rm H\textsc{i}}} \times 10^{0.4A_{\rm UV}},
\end{equation}
where $\mathcal{R_{\rm int}}$ is the intrinsic LyC to non-ionizing UV flux ratio, $e^{-\tau_{\lambda}^{\rm H\textsc{i}}}$ is the line-of-sight transmission through the IGM and CGM, and $A_{\rm UV}$ is the UV dust attenuation\footnote{As measured at the rest-frame effective wavelength of 
the F606W filter, which is typically $\simeq 1300$\,\AA\,for our sample.}.

While $\mathcal{R_{\rm int}}$ and $A_{\rm UV}$ can be estimated using standard stellar population fitting and/or empirical methods, the crucial complicating factor is the value of $e^{-\tau_{\lambda}^{\rm H\textsc{i}}}$, which is strongly sight-line dependent and can take a range of values drawn from a highly non-Gaussian distribution \citep[e.g.][]{steidel+18}.
As a result, there is no unique mapping between $f_{\rm esc}$ and $\mathcal{R_{\rm obs}}$ for individual objects, and one must account for the full distribution of possible $\mathcal{R_{\rm obs}}$ values at a given $f_{\rm esc}$.
In this section we describe in detail how our model for $\mathcal{R_{\rm obs}}(f_{\rm esc})$ was constructed, focusing on each of the three key quantities ($\mathcal{R_{\rm int}}$, $e^{-\tau_{\lambda}^{\rm H\textsc{i}}}$, $A_{\rm UV}$) in turn.

\subsection{Intrinsic LyC to non-ionizing UV flux ratio}\label{subsec:intrinsic_sed}

The intrinsic LyC to non-ionizing UV flux ratio is determined by the properties of the underlying stellar population, which sets the shape of the intrinsic SED, and the galaxy redshift, which fixes the specific rest-frame wavelength regions covered by the $U$ and $V_{606}-$band filters. 

In order to define an intrinsic SED that is representative of the average properties of our sample, we first constructed a
stack of the VANDELS spectra, following the method described in \citet{cullen+19}. 
The best-fitting stellar metallicity of this stacked spectrum was then determined by fitting the Binary Population and Spectra Synthesis version 2.2 (BPASSv2.2) SPS models \citep{eldridge+17, stanway+18} following the full spectral fitting approach outlined in \citet{cullen+19}, assuming a constant star-formation history over a 100 Myr timescale, binary stellar evolution, and a standard \citet{kroupa+01} IMF with an upper mass limit of 100$\rm{M}_{\odot}$.
Our choice of the BPASS models was motivated by observations that suggest these models yield the best predictions for the ionizing continuum spectra of high-redshift stellar populations \citep[e.g.][]{steidel+16,reddy+21}.

The best-fitting BPASS model had a metallicity of $Z_{\star}\simeq0.001$ $\simeq 0.07$ $\rm{Z}_{\odot}$ \citep[assuming][]{asplund+09}, consistent with previous estimates of the average stellar metallicity of galaxies at similar redshifts and stellar masses \citep{cullen+19, cullen+21, kashino+21}. 
This best-fitting model was adopted as the representative intrinsic SED for our sample, and the individual $\mathcal{R_{\rm int}}$ values were then calculated by integrating through the $U$ and $V_{606}-$band filters 
at the redshift of each galaxy. Across our final galaxy sample the individual values of $\mathcal{R_{\rm int}}$ range from $0.17$ to $0.20$ with a median value of $\mathcal{R_{\rm int}}=0.19$.
It is worth nothing that this is essentially the same intrinsic SED used to infer $\langle f_{\rm{esc}}\rangle$ and other related parameters in the recent the KLCS spectroscopic analyses at $z\sim3$ \citep[e.g.][]{steidel+18,pahl+21}.

\subsection{IGM and CGM transmission}\label{subsec:igmcgm_distributions}

At the redshift of our sample, the nuisance parameter with the largest influence on the derived value of $\langle f_{\rm esc}\rangle$ is the optical depth of the IGM and CGM, which determines the transmitted fraction of ionizing photons through the intervening \hi \ along the line-of-sight to each galaxy ($e^{-\tau_{\lambda}^{\rm H\textsc{i}}}$). The optical depth can vary significantly with sight line, depending on the exact distribution of neutral clouds (as a function of column density and redshift), and therefore must be accounted for in a probabilistic sense.

In many previous studies, it has been common to only consider the contribution of the IGM when accounting for \hi \ optical depth \citep[e.g][]{vanzella+10a}.
However, as galaxies exist in regions of gas overdensities, and are known to be surrounded by significant quantities of \hi \ in their CGM \citep[out to $\simeq 700$ physical kpc; e.g.][]{rudie+12,rudie+13}, sight lines to galaxies are not representative of random sight lines through the Universe.
As a result, it is more accurate to account for \emph{both} the IGM and CGM when considering the optical depth of \hi \ towards galaxies \citep[e.g.][]{steidel+18, pahl+21}.

To account for the IGM and CGM contribution we generated transmission curves ($e^{-\tau_{\lambda}^{\rm H\textsc{i}}}$) using the parameterization for the column density and redshift distribution of \hi \ clouds given in \citet{steidel+18}. Specifically, we generated $10,000$ individual sight lines in six separate redshift bins ($z=3.4,3.5,3.6,3.7,3.8,3.9$), covering the full redshift range of our sample. 
Full details of the method used to generate the individual sight lines are provided in Appendix \ref{app:igm_model}, and examples of nine random sight lines at $z=3.4$ are shown in the left-hand panel of Fig. \ref{fig:igm_cgm_transmission}, highlighting the significant variation.

For a given galaxy, a value of $e^{-\tau_{\lambda}^{\rm H\textsc{i}}}$ is obtained by selecting a random sight line at the nearest redshift and integrating through the $U$-band filter.
The right-hand panel of Fig. \ref{fig:igm_cgm_transmission} illustrates how the resulting distribution of $e^{-\tau_{\lambda}^{\rm H\textsc{i}}}$ is strongly peaked at zero, with a highly non-Gaussian shape.

\subsection{UV dust attenuation}\label{subsec:auv}

After the IGM+CGM transmission, the model parameter that has the largest systematic influence on the derived value of $\langle f_{\rm esc}\rangle $ is the UV dust attenuation.
In this study, we have taken advantage of the rest-frame UV VANDELS spectra to  adopt an empirical approach to calculating the level of UV attenuation on a galaxy-by-galaxy basis.
By comparing to our adopted intrinsic SED model (see §\ref{subsec:intrinsic_sed}), which has a UV spectral slope of $\beta_{\rm int}=-2.44\pm{0.02}$, we calculated the 
value of $A_{\rm UV}$ for each object by measuring the observed UV spectral slope ($\beta_{\rm obs}$) from its VANDELS spectrum.

The first step in this process was fitting a power-law ($f_{\lambda}~\propto~\lambda^{\beta}$) to each of the VANDELS spectra
over the wavelength range $1300~-~1800$~\AA, within the continuum windows specified by \citet{calzetti+94}. 
Fitting $\beta$ in this fashion, the final galaxy sample has $\langle \beta_{\rm spec} \rangle = -1.26\pm{0.04}$, with a median value of $\beta_{\rm spec} = -1.29$.
The average of these individual $\beta$ estimates is fully consistent with the value derived from the stacked spectrum of the full sample ($\beta_{\rm stack}=-1.24\pm{0.03}$).

Armed with the individual values of $\beta_{\rm obs}$, it was then possible to
calculate individual determinations of $A_{1600}$ based on the value of $\Delta \beta=\beta_{\rm obs}-\beta_{\rm int}$. 
However, making this conversion requires a decision to be made on the form of the UV attenuation curve. 
Unfortunately, the average form of the UV attenuation curve at high redshift is still a matter of debate, with no consensus having been reached in the literature \cite[e.g.][]{cullen+18, mclure+18, reddy+18, shivaei+20}, and we therefore chose to employ a dust curve that is at least consistent with both the VANDELS spectra and our choice of intrinsic SED model.
To do this, we fitted the stacked VANDELS spectrum using the BPASS intrinsic SED model, attenuated by a dust curve parameterized following \citet{salim+18}. 
According to this formulation, the \citet{calzetti+00} attenuation curve is modified by a power-law exponent ($\delta$), such that $\delta=0.0$ corresponds to the Calzetti starburst curve and $\delta \simeq -0.5$ is close to the SMC extinction curve \citep[e.g.][]{gordon+03}. Fitting the stacked spectrum over the wavelength range $1300-1800$ \AA\, returned a best-fitting dust slope of $\delta=-0.25^{+0.37}_{-0.27}$, with no $2175$ \AA\, dust bump. 

Based on this dust curve, we proceeded to convert the individual values of $\Delta \beta$ measured for each galaxy into attenuation at 1600 \AA, using the relation: $A_{1600}=1.28\times \Delta\beta$. 
For each object, we then calculated the value of $A_{\rm UV}$ at the effective wavelength of the $V_{606}$ filter using $A_{\rm UV}\simeq 1.2\times A_{1600}$, where the constant has a slight redshift dependence.
The error on $A_{\rm UV}$ ($\sigma_{A_{\rm UV}}$) was estimated by propagating the error on $\Delta \beta$.

\begin{figure*}
        \centerline{\includegraphics[width=2\columnwidth]{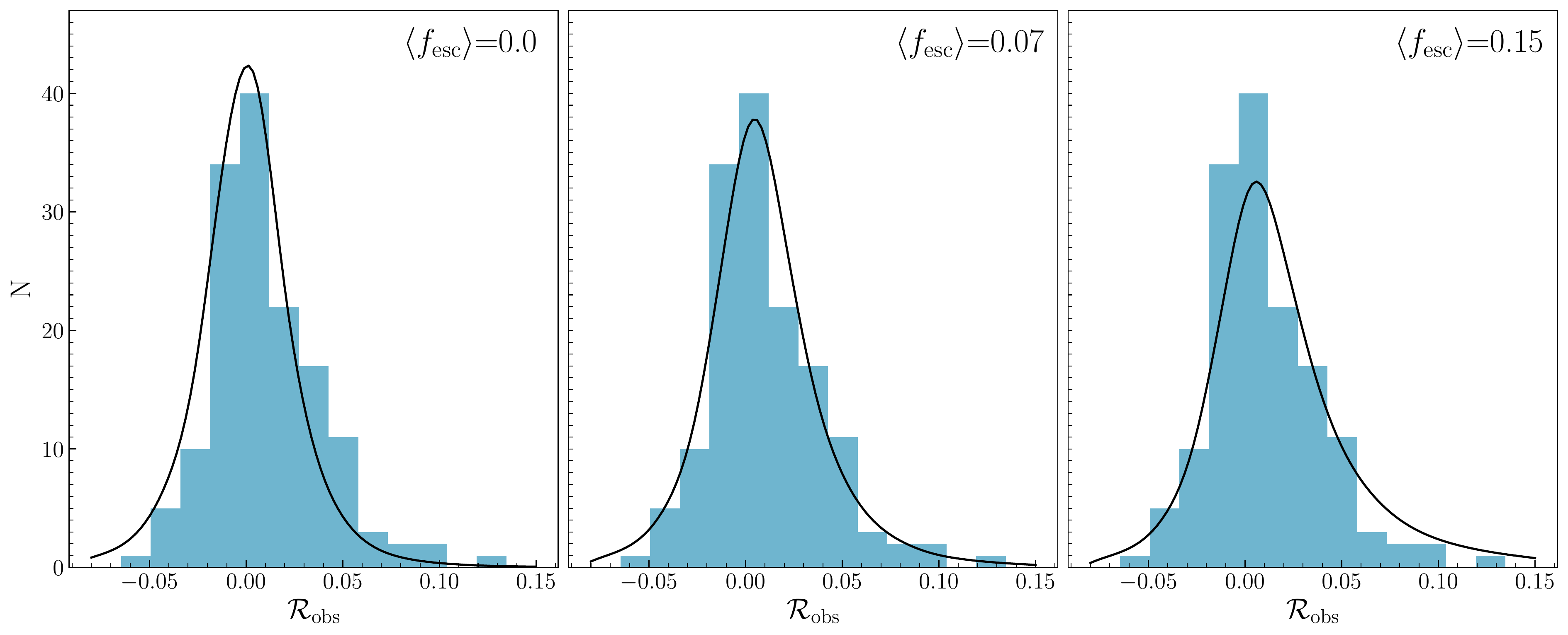}}
        \caption{A comparison between the $\mathcal{R_{\rm{obs}}}$ distribution for our full sample of 148 star-forming galaxies (blue) and the output of the model described in Section \ref{sec:model} (black line), for three different values of $\langle f_{\rm esc}\rangle$. The best-fitting model returned by the maximum likelihood method described in Section \ref{sec:distn_method} is shown in the middle panel, and corresponds to $\langle f_{\rm esc}\rangle=0.07\pm0.02$. It can be seen that the model predictions for $\langle f_{\rm esc}\rangle=0$ and $\langle f_{\rm esc}\rangle=0.15$ are not compatible with observed data, and both 
        are formally excluded at $\geq 3\sigma$ (see Fig. \ref{fig:chi2_fullsample}).}
        \label{fig:distngrid}
\end{figure*}
\subsection{Constructing model $\bmath {\mathcal{R}_{\rm obs}(f_{\rm esc})}$ distributions}\label{subsec:fesc_model_distributions}

Combining these three components, it is possible to construct the expected distribution of $\mathcal{R_{\rm obs}}$ as a function of $f_{\rm esc}$.
The model distribution can then be statistically compared to the observed $\mathcal{R_{\rm obs}}$ distribution, to place constraints on $\langle f_{\rm esc}\rangle$ for any given sample.
We adopted a Monte Carlo procedure for producing model $\mathcal{R_{\rm obs}}$ distributions as a function of $f_{\rm esc}$. For a set of $\mathcal{N}$ galaxies drawn from the full galaxy sample, we 
performed the following steps:
\begin{enumerate}
\item For each galaxy, select a random IGM+CGM sight line at the appropriate redshift and calculate the value of $e^{-\tau_{\lambda}^{\rm H\textsc{i}}}$ integrated through the $U-$ band filter.
\item Set the value of $A_{\rm UV}$ by perturbing the measured value assuming a Gaussian scatter of $\sigma_{A_{\rm UV}}$ and ensuring that $A_{\rm UV} \geq 0$.
\item Using these two values, and the adopted value of $\mathcal{R_{\rm int}}$, calculate the expected $\mathcal{R_{\rm obs}}$ using Equation \ref{eq:fesc_eq}.
\item Perturb $\mathcal{R_{\rm obs}}$ according to the depth of the $U$ and $V_{606}-$ band mosaics at the position of the galaxy.
\end{enumerate}
These steps yield $\mathcal{N}$ model $\mathcal{R_{\rm obs}}$ values. The process was then repeated $10,000$ times in order to build-up the average model distribution that could then
be directly compared to the observed data, as shown in Fig.~\ref{fig:distngrid}.

\section{Results}\label{sec:results}
In this section we describe how we estimated $\langle f_{\rm esc} \rangle$ for our final galaxy sample, using two approaches: (i) a binned maximum likelihood method and (ii) a Bayesian framework for combining individual $f_{\rm esc}$ estimates.
We also explore whether trends in $\langle f_{\rm esc} \rangle$ can be identified by splitting 
our sample on the basis of properties that are expected to correlate with $f_{\rm esc}$; such as the equivalent width of Ly$\alpha$, UV continuum slope $\beta$ (a proxy for dust attenuation), galaxy stellar mass and intrinsic UV luminosity.

 \begin{figure}
         \centerline{\includegraphics[width=\columnwidth]{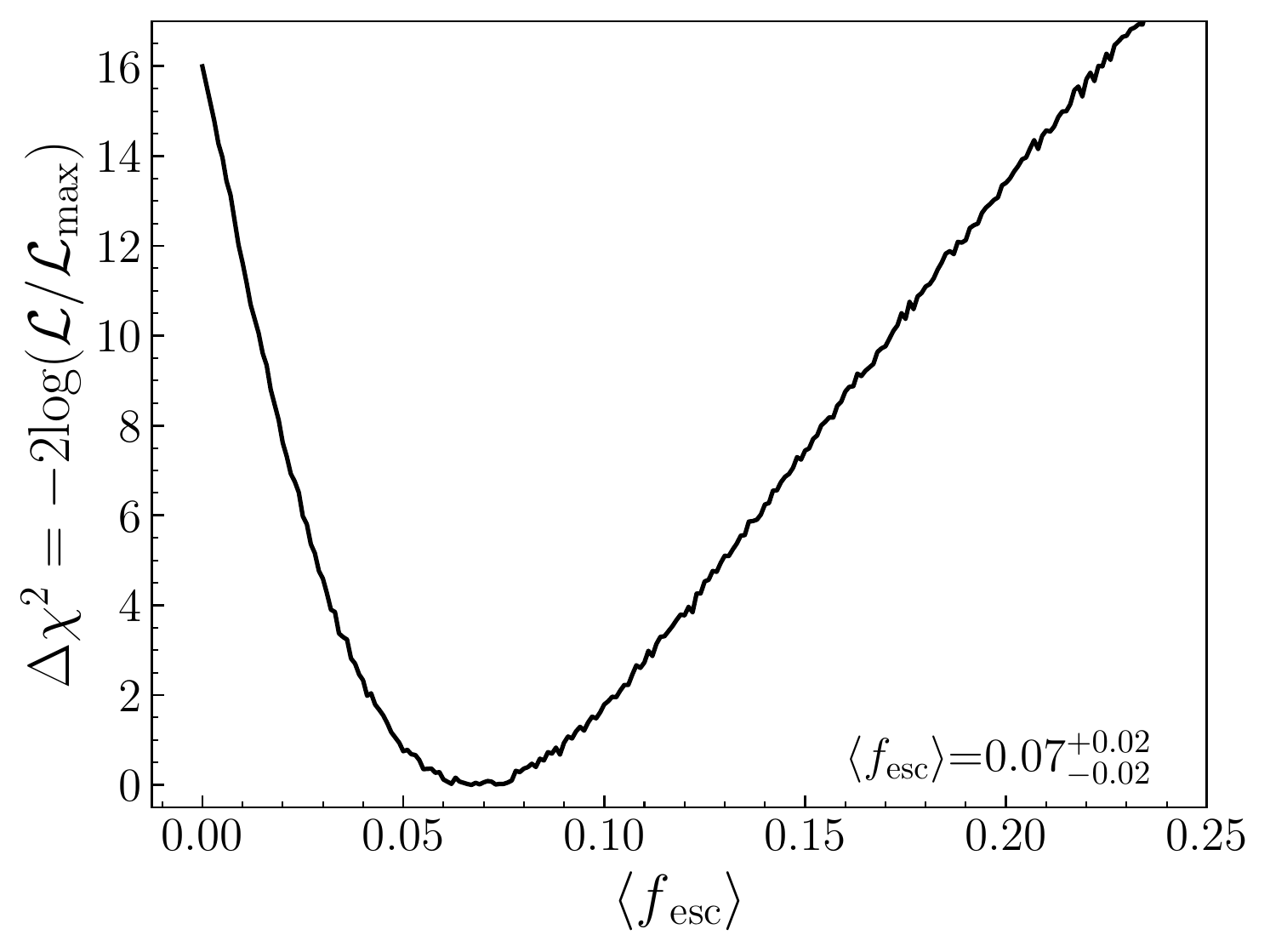}}
         \caption{The result of the maximum likelihood fit to the $\mathcal{R_{\rm obs}}$ distribution of the full sample of 148 star-forming galaxies. The best-fitting value is found to be $\langle f_{\rm esc} \rangle = 0.07\pm0.02$ and $\langle f_{\rm esc} \rangle=0.0$ is excluded at $\geq 3\sigma$ confidence (i.e. $\Delta \chi^2\geq 9$).}
         \label{fig:chi2_fullsample}
     \end{figure}

\subsection{Maximum Likelihood}\label{sec:distn_method}

The maximum likelihood technique is based upon a comparison between the observed and model $\mathcal{R_{\rm obs}}$ distributions. 
Model $\mathcal{R_{\rm obs}}$ distributions were built for a grid of $\langle f_{\rm esc} \rangle$ values between $0$ and $1$ with an interval of $\Delta \langle f_{\rm esc} \rangle =0.001$, following the procedure outlined in Section 
\ref{sec:model}.
The fitting procedure was to maximize the following log-likelihood function:
\begin{equation}
    \ln{\mathcal{L}} = \textstyle \sum_{i} n_{i} \ln{\ p_{i}},
\end{equation}
where the summation runs over the $i$ bins of the $\mathcal{R_{\rm obs}}$ histogram (see Fig. \ref{fig:observedratio}), $n_{i}$ is the number of galaxies in bin $i$ and $p_i$ is the probability of finding a galaxy within bin $i$ for a given value of $f_{\rm esc}$.
The probability $p_{i}$ is naturally defined as $n_{i}^{m}/N$, where $n_{i}^{m}$ is the number of galaxies in bin $i$ predicted by the model, for a given $f_{\rm esc}$, and $N$ is the total number of galaxies in the sample ($N=148$). The $1 \sigma$ confidence interval can be estimated via:
\begin{equation}
\Delta \chi^{2} = -2\ln{\left(\mathcal{L}/\mathcal{L}_{\mathrm{max}}\right)} =1,
\end{equation}
where $\mathcal{L}_{\mathrm{max}}$ is the maximum likelihood value.
Applying this technique we find clear evidence for a non-zero $\langle f_{\rm esc}\rangle$ at the $> 3\sigma$ level, with a best-fitting value of $\langle f_{\rm esc} \rangle = 0.07\pm0.02$ (see Fig.~\ref{fig:chi2_fullsample}).

Fig. \ref{fig:distngrid} provides a visual illustration of this result. It can be seen from the middle panel that the model distribution corresponding to $\langle f_{\rm esc} \rangle = 0.07$ provides an excellent description of the data. In contrast, the $\langle f_{\rm esc} \rangle = 0$ model predicts too many galaxies with $\mathcal{R_{\rm obs}}~<~0$, while the $\langle f_{\rm esc} \rangle = 0.15$ model predicts a positive tail in excess of what is observed and underestimates the $\mathcal{R_{\rm obs}}=0$ peak. 

\subsection{Bayesian Inference}\label{sec:bayes_method}

To complement the maximum likelihood fitting, we adopted a second approach for estimating $\langle f_{\rm esc} \rangle$ that utilises the individual posterior probabilities 
for $f_{\rm esc}$ of each galaxy. Using Bayes' theorem, the posterior 
probability for $f_{\rm esc}$ is given by:
\begin{equation}
p(f_{\rm esc} | \mathcal{R}_{\rm obs}) \propto\int p(\mathcal{R}_{\rm obs}| f_{\rm esc}, \Theta)p(f_{\rm esc})p(\Theta)d\Theta
\end{equation}
where $\Theta=(\mathcal{R}_{\rm int}, e^{-\tau_{\rm U}^{\rm H\textsc{i}}}, A_{\rm UV})$ and $p(\mathcal{R}_{\rm obs}| f_{\rm esc}, \Theta)$ is the likelihood for $R_{\rm obs}$, $p(f_{\rm esc})$ is the prior on the escape fraction, and $p(\Theta)$ is the prior on the additional free parameters. For a given galaxy, we can write the log-likelihood as, 
\begin{equation}
    \mathrm{ln} \ p(\mathcal{R}_{\rm obs}| f_{\rm esc}, \Theta) =  \frac{-(\mathcal{R}(f_{\rm esc},\Theta)-\mathcal{R}_{\rm obs})^2}{2 \sigma_{\rm obs}^2} - \rm{ln} (\sqrt{2\pi}\sigma_{\rm obs}),
\end{equation}
where $\sigma_{\rm obs}$ is the error on $\mathcal{R}_{\rm obs}$ and $\mathcal{R}(f_{\rm esc},\Theta)$ is the predicted value of $\mathcal{R}$ for a given set of 
input parameters, according to Equation \ref{eq:fesc_eq}.

\begin{figure}
         \centerline{\includegraphics[width=\columnwidth]{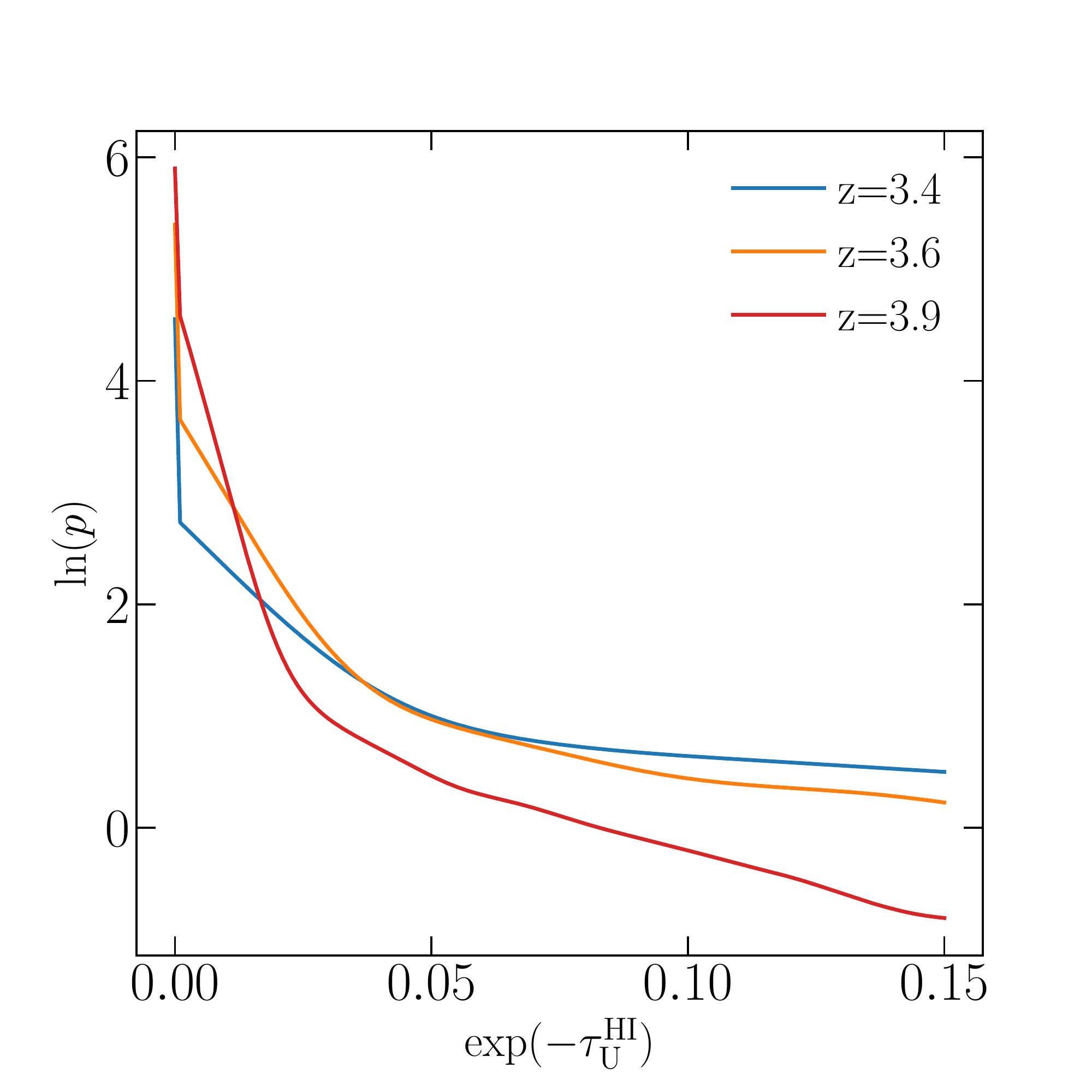}}
         \caption{Probability density functions for $e^{-\tau_{\rm U}^{\rm H\textsc{i}}}$, used as priors in our Bayesian inference methodology (Section \ref{sec:bayes_method}).
         To generate the smooth probability density functions we performed a kernel density estimation of the distribution of 10,000 sight lines at each redshift (examples of these can be seen in Fig. \ref{fig:igm_cgm_transmission}).
         The increase in probability density at low transmission (i.e., high IGM+CGM optical depth) with increasing redshift can clearly be seen.}
         \label{fig:igmcgm_prior}
     \end{figure}

The crucial aspect of the Bayesian approach then becomes determining the priors for each parameter.
For simplicity, we assumed a fixed value for $\mathcal{R}_{\rm int}$ for each galaxy (see Section \ref{subsec:intrinsic_sed}). We adopted the following priors for each of the other three free parameters: (i) for $f_{\rm esc}$ we assume a uniform prior between $0$ and $1$; (ii) for $A_{\rm UV}$ we assume a Gaussian prior with mean and standard deviation given by the individual fits to the UV continuum slope of each galaxy (Section \ref{subsec:auv}); (iii) finally, to generate a prior on $e^{-\tau_{\rm U}^{\rm H\textsc{i}}}$ 
we perform a kernel density estimation to turn the distribution of sight lines (see Section \ref{subsec:igmcgm_distributions}) into a smooth probability distribution. The resulting prior distributions for $e^{-\tau_{\rm U}^{\rm H\textsc{i}}}$ as a function of redshift are shown in Fig. \ref{fig:igmcgm_prior}.

\begin{figure*}
        \begin{subfigure}[b]{1\columnwidth}
            \includegraphics[width=\textwidth]{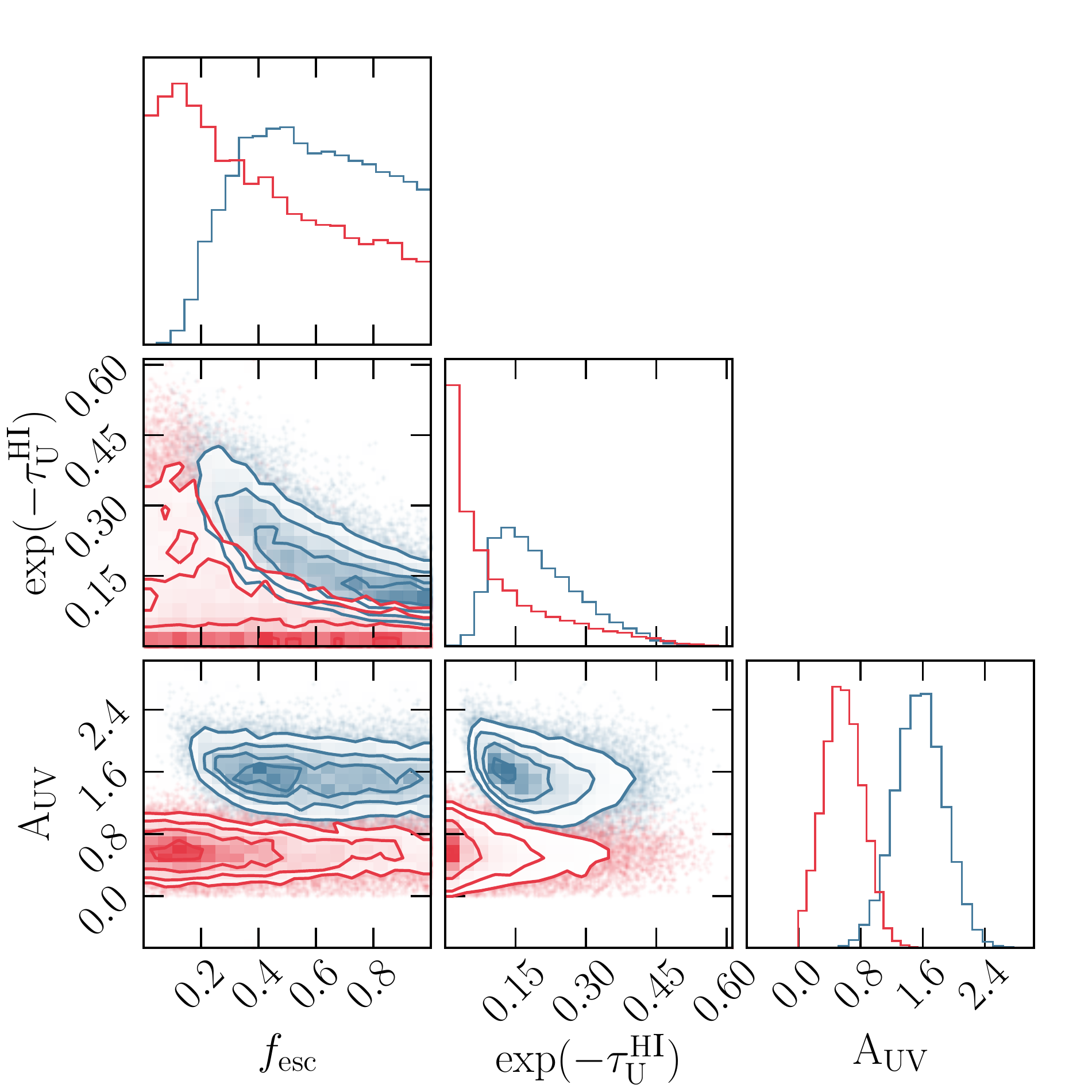}
        \end{subfigure}
        \hfill
        \begin{subfigure}[b]{1\columnwidth}
            \includegraphics[width=\textwidth]{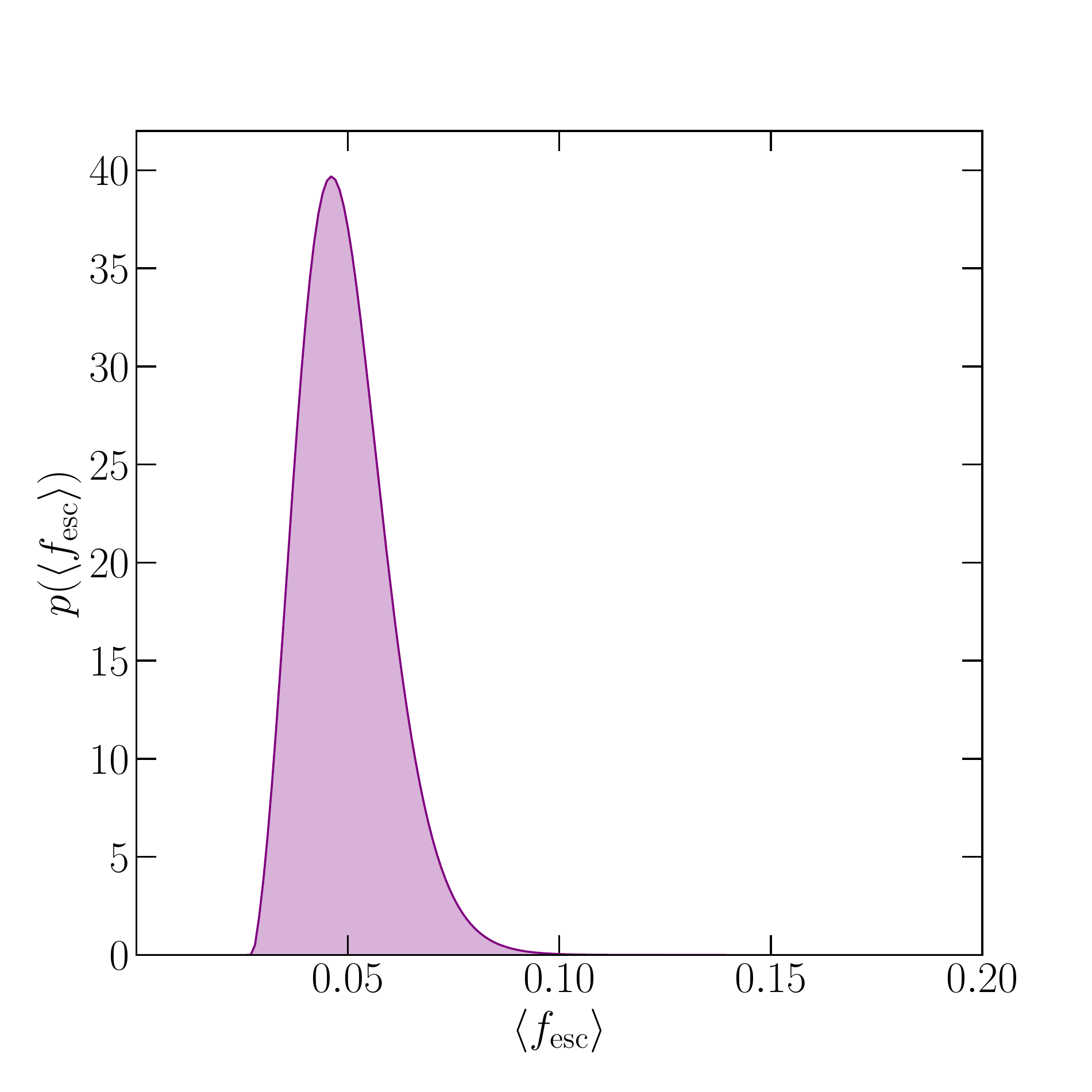}
        \end{subfigure}
         \caption{\emph{Left}: A corner plot showing the 1D and 2D marginalised posteriors for $f_{\rm esc}$, $A_{\rm UV}$ and $e^{-\tau_{\rm U}^{\rm H\textsc{i}}}$ for two example galaxies drawn from our final sample.
         The blue posteriors show one of the two galaxies in our full sample with a robust ($\geq 5 \sigma$) individual $\mathcal{R}_{\rm obs}$ detection.
         The red posteriors show an example of a typical galaxy that is not individually detected.
         To calculate the average $\langle f_{\rm esc}\rangle $ for a given galaxy sample, the individual $f_{\rm esc}$ posteriors (i.e., curves in the upper left panel) were multiplied together. \emph{Right}: The posterior probability $p(\langle f_{\rm esc} \rangle | \{ R_{\rm obs} \})$ for the full galaxy sample, corresponding to $\langle f_{\rm esc} \rangle = 0.05\pm0.01$.}
         \label{fig:bayes_result}
     \end{figure*}

Armed with the likelihood and prior distributions we determined the posterior of $f_{\rm esc}$ for each galaxy using an MCMC sampling technique \citep{foreman-mackey+2013}. 
Examples of the posterior distributions for a detected (S/N$\geq 5\sigma$) and non-detected galaxy in the $U-$ band are shown in the left-hand panel of Fig.~\ref{fig:bayes_result}. The value of $\langle f_{\rm esc} \rangle$ can then be obtained by multiplying the individual posteriors:
\begin{equation}\label{eq:multi_pfesc}
p(\langle f_{\rm esc} \rangle | \{ R_{\rm obs} \}) \propto \textstyle \prod^{N_{\rm gal}}_{i} p_{i}(f_{\rm esc}).
\end{equation}

In the right-hand panel of Fig.~\ref{fig:bayes_result} we show the resulting $p(\langle f_{\rm esc} \rangle | \{ R_{\rm obs} \})$ for our full sample.
It is important to note that equation \ref{eq:multi_pfesc} is only valid under that assumption that each galaxy has the same value of $f_{\rm esc}$.
Therefore, $\langle f_{\rm esc} \rangle$ should be interpreted as the most likely value of $f_{\rm esc}$ assuming a uniform value across the sample, rather than the average of $148$ individual (potentially different) values.
In this way, $\langle f_{\rm esc} \rangle$ has the same physical interpretation as $\langle f_{\rm esc} \rangle$ derived from the maximum likelihood fitting above.
Using this method, we infer a best-fitting value of $\langle f_{\rm esc} \rangle = 0.05\pm0.01$, fully consistent (within $1\sigma$) with the value inferred from the maximum likelihood approach, and inconsistent with $\langle f_{\rm esc} \rangle = 0$ at the $\geq 3\sigma$ level.

\subsection{Escape fraction dependence on galaxy physical properties}\label{subsec:sample_splits}
Having established that $\langle f_{\rm esc}\rangle$ for the full sample is non-zero, it is clearly of interest to investigate whether
or not there is any indication that $f_{\rm esc}$ correlates with other galaxy properties. As discussed in the introduction, reliable 
proxies for LyC escape are required in order to infer $f_{\rm{esc}}$ during the reionization era, where direct measurements are not possible.

One of the most prominent amongst potential $f_{\rm{esc}}$ indicators is the equivalent width of the \Lya line (\wlya), which is expected to correlate with LyC escape since both are sensitive to the column density and distribution of \hi \ within galaxies \citep[e.g.][]{verhamme+15, gronke+15, dijkstra+16}. Observational evidence in support of this connection has recently been reported via spectroscopic analyses of galaxies at $z\simeq3$, which find strong evidence for a correlation between $f_{\rm esc}$ and \wlya \ \citep{marchi+17,steidel+18,pahl+21}, as well as \wlya \ and the covering fraction of \hi \ \citep[][]{reddy+16, gazanges2020, reddy+21, saldana-lopez22}.
Other likely indirect tracers of the neutral \hi \ column density of galaxies include the dust content (traced by the UV continuum slope, $\beta$) and stellar mass ($M_{\star}$); indeed, both of these quantities are known to be linked to the escape of \Lya photons \citep{du+18,cullen+20}. 

Finally, it is also of interest to investigate whether $f_{\rm esc}$ and UV luminosity are correlated, given that calculating  the global ionizing background during the EoR typically relies on integrating down the UV galaxy luminosity function with an assumption that $f_{\rm esc}$ is constant \citep[e.g.][]{robertson+15}. Below we explore the correlation between $f_{\rm esc}$ and each of these galaxy properties (\wlya, $\beta$, $M_{\star}$, $L_{\rm UV}$), in turn.

\subsubsection{\wlya}
\begin{figure*}
        \centerline{\includegraphics[width=14cm]{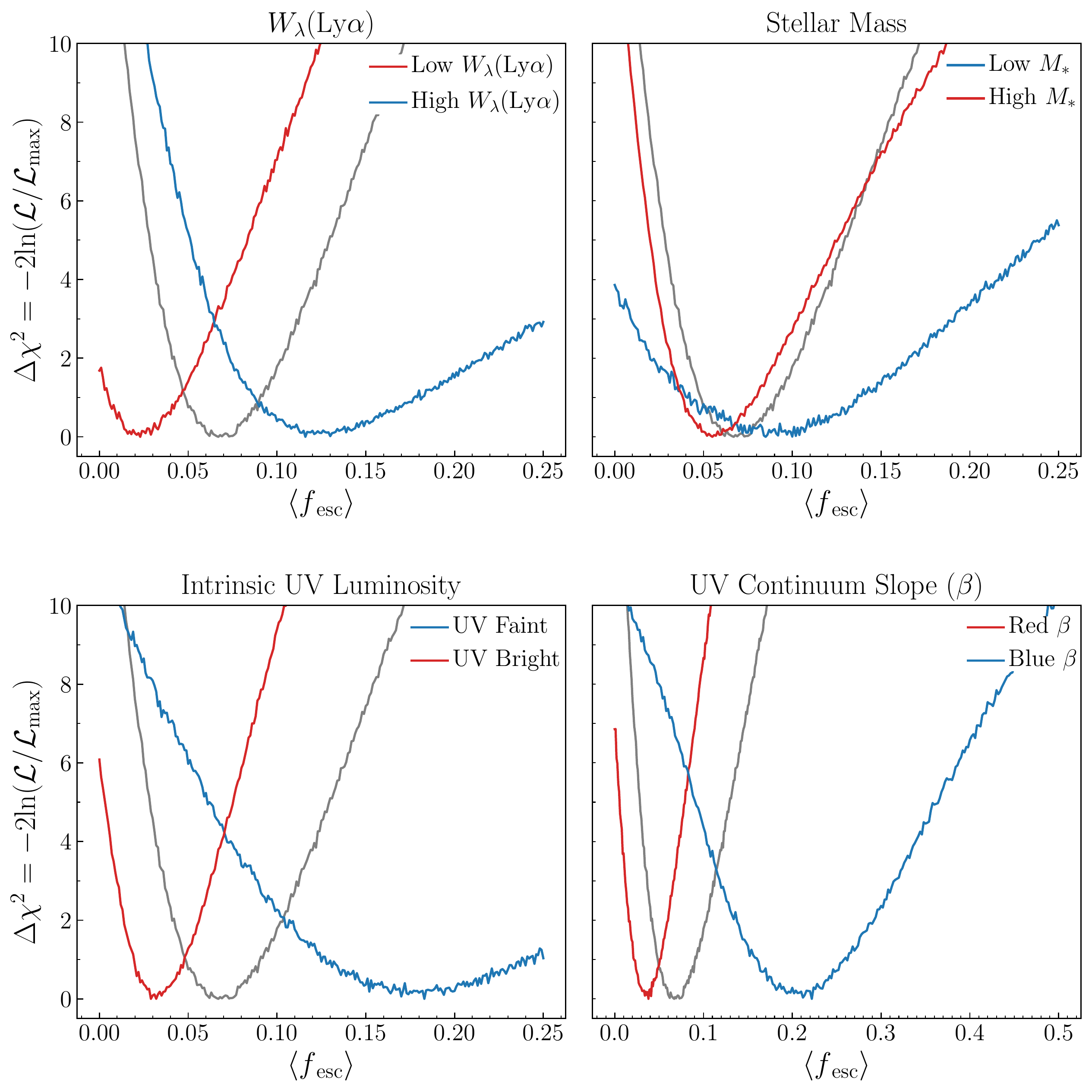}}        \caption{The constraints on $\langle f_{\rm esc}\rangle$ returned by the maximum-likelihood technique for the four sample splits described in Section \ref{subsec:sample_splits}. In each panel the red and blue curves show the $\langle f_{\rm esc}\rangle$ constraints when the full sample is split in two at the median value of the physical parameter in question. The red and blue curves can be identified by the labels in the top-right corner of each panel. For reference, the grey curve in each panel shows the constraint on $\langle f_{\rm esc}\rangle$ for the full galaxy sample.}
        \label{fig:sampleChi2}
\end{figure*}

\begin{figure*}
    \centerline{\includegraphics[width=16cm]{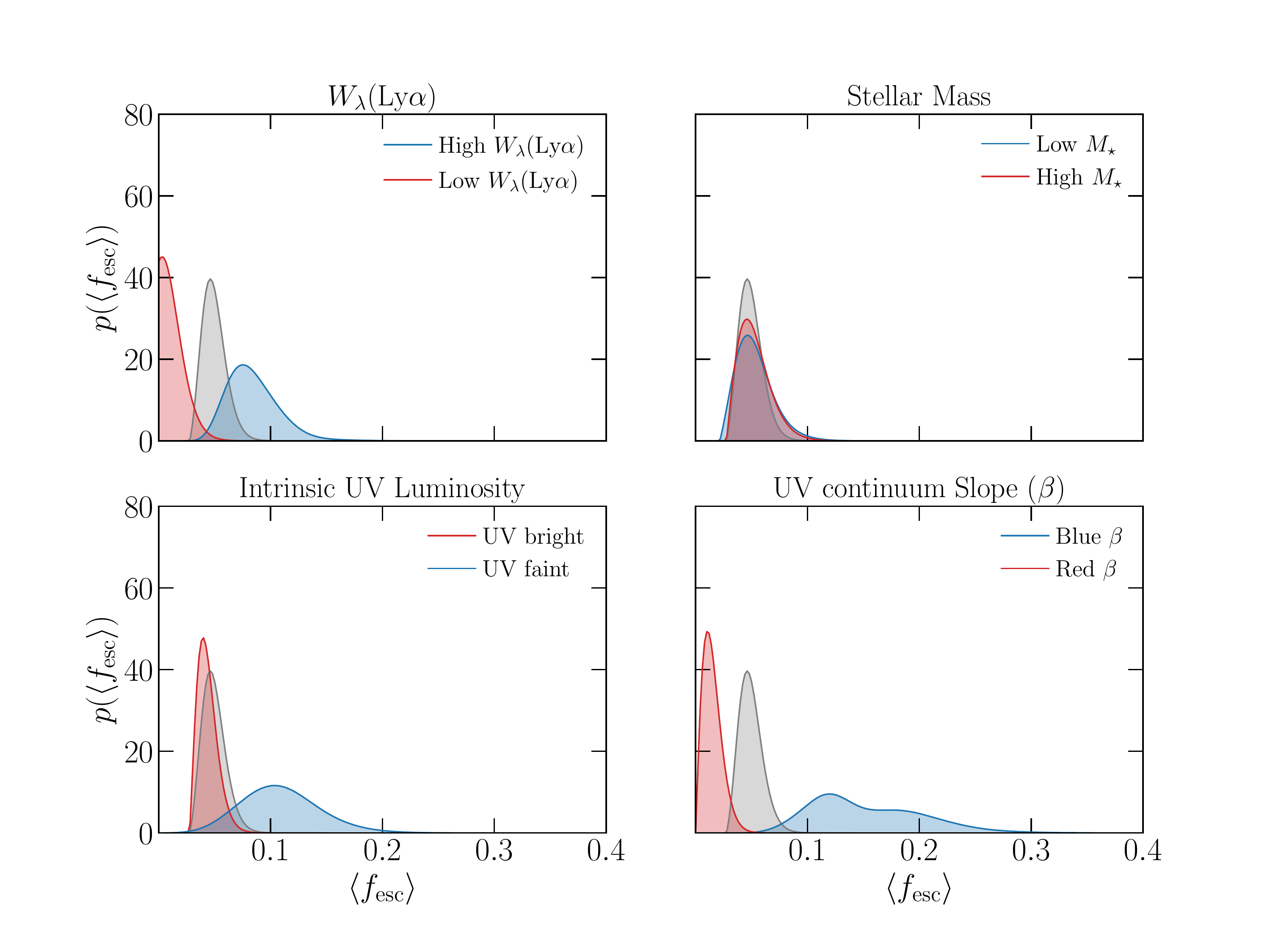}}
    \caption{The constraints on $\langle f_{\rm esc}\rangle$ returned by the Bayesian inference technique for the four sample splits described in Section \ref{subsec:sample_splits}, in the same format as Fig.~\ref{fig:sampleChi2}. The red and blue curves show the posterior probability distributions for the sample splits, whereas
    the grey curve in each panel shows the posterior probability distribution of $\langle f_{\rm esc}\rangle$ for the full sample. As before, the red and blue curves can be identified by the labels in the top-right corner of each panel.}
    \label{fig:bayes_subsamples}
\end{figure*}

To investigate the link between $f_{\rm esc}$ and \wlya, we split the full sample in half at the median value of \wlya=$-6$\,\AA.
After excluding two galaxies with unreliable \wlya \ measurements due to artefacts in the VANDELS spectra, the 
resulting low$-$ and high$-$\wlya \ sub-samples had median equivalent widths of \wlya$=-14.2$\,\AA\, and \wlya$=4.9$\,\AA, respectively. Fitting the two sub-samples using the maximum likelihood technique returned best-fitting values of 
$\langle f_{\rm esc}\rangle =0.02^{+0.02}_{-0.02}$ for the low$-$\wlya \ sub-sample and $\langle f_{\rm esc}\rangle~=~0.12^{+0.06}_{-0.04}$ for the high$-$\wlya\, sub-sample (Fig. \ref{fig:sampleChi2}).
Using the Bayesian inference methodology, the constraints for the low and high$-$\wlya \ sub-samples were $\langle f_{\rm esc}\rangle< 0.03  \ (2\sigma)$ and $\langle f_{\rm esc}\rangle =0.08^{+0.02}_{-0.02}$ (Fig. \ref{fig:bayes_subsamples}).
These results represent significant evidence ($>3 \sigma$) for an increase in $f_{\rm esc}$ with increasing \wlya, in broad agreement with recent spectroscopic studies \citep{steidel+18, pahl+21}.

\renewcommand{\arraystretch}{1.3}
\begin{table}
    \centering
    \caption{Summary of the best-fitting $\langle f_{\rm esc} \rangle$ values of the various samples discussed in Section \ref{sec:results}. Results are quoted for both the maximum likelihood and Bayesian inferences methods. Upper limits represent $2\sigma$ constraints.}\label{table:fesc_constrains}
    \begin{tabular}{lcc}
    \hline
        Sample & Maximum Likelihood &  Bayesian Inference \\
        \hline
        \hline
        Full Sample & $0.07^{+0.02}_{-0.02}$ & $0.05^{+0.01}_{-0.01}$ \\
        \hline
        Low \wlya & $0.02^{+0.02}_{-0.02}$ & $<0.03$ \\
        High \wlya & $0.12^{+0.06}_{-0.04}$ & $0.08^{+0.02}_{-0.02}$ \\
        \hline
        Red $\beta$ & $0.04^{+0.01}_{-0.02}$ & $<0.03$ \\
        Blue $\beta$ & $0.22^{+0.04}_{-0.06}$ & $0.14^{+0.06}_{-0.04}$ \\
        \hline
        High-$M_{\star}$ & $0.06^{+0.02}_{-0.02}$ & $0.05^{+0.02}_{-0.01}$ \\
        Low-$M_{\star}$ & $0.09^{+0.05}_{-0.04}$ & $0.05^{+0.02}_{-0.01}$ \\
        \hline
        Intrinsic UV-bright & $0.03^{+0.02}_{-0.01}$ & $0.04^{+0.01}_{-0.01}$ \\
        Intrinsic UV-faint & $0.18^{+0.06}_{-0.05}$ & $0.10^{+0.04}_{-0.03}$ \\
    \hline
    \end{tabular}
    
\end{table}
\subsubsection{UV continuum slope}
Next, we looked for a link between $f_{\rm esc}$ and the observed UV continuum slope $\beta_{\rm obs}$ (a proxy 
for dust attenuation at UV wavelengths). 
A $f_{\rm esc}-\beta_{\rm obs}$ correlation is expected due to the sensitivity of LyC photon escape to the dust and \hi \ column density of the ISM. 

Again, the full galaxy sample was split in half at the median value of $\beta_{\rm obs}=-1.29$, producing low attenuation (`blue', with median $\beta_{\rm obs} = -1.62$) and high attenuation (`red', with median $\beta_{\rm obs} = -0.92$) sub-samples. 
Applying the maximum likelihood approach returned best-fitting values of  $\langle f_{\rm esc}\rangle =0.22^{+0.04}_{-0.06}$ and $\langle f_{\rm esc}\rangle =0.04^{+0.01}_{-0.02}$ for the low- and high-attenuation sub-samples, respectively (Fig. \ref{fig:sampleChi2}). 
Similarly, the Bayesian inference approach returned $\langle f_{\rm esc}\rangle =0.14^{+0.06}_{-0.04}$ for the low-attenuation sub-sample and $\langle f_{\rm esc}\rangle < 0.03 \ (2\sigma)$ for the high-attenuation sub-sample (Fig. \ref{fig:bayes_subsamples}).
Taken together, these results again represent significant evidence in favour of a picture in which galaxies with lower levels of UV dust attenuation display higher values of $\langle f_{\rm esc}\rangle$.

\subsubsection{Stellar mass}
Although \wlya \ and $\beta_{\rm obs}$ are two galaxy properties with a clear and direct link to $f_{\rm esc}$, it is also interesting to
examine any correlation between $f_{\rm esc}$ and stellar mass, a property that is already know to correlate with both \wlya \ and UV dust attenuation \citep[e.g][]{mclure+18, cullen+20}. 
Splitting the sample into low-$M_{\star}$ (median $=10^{8.72}\mathrm{M}_{\odot}$) and high-$M_{\star}$ (median $=10^{9.22}\mathrm{M}_{\odot}$) sub-samples, the maximum likelihood approach returned best-fitting values of 
$\langle f_{\rm esc}\rangle=0.09^{+0.05}_{-0.04}$ and $\langle f_{\rm esc}\rangle =0.06^{+0.02}_{-0.02}$, respectively (Fig. \ref{fig:sampleChi2}).
With our Bayesian inference approach, we derive corresponding constraints of $\langle f_{\rm esc}\rangle=0.05^{+0.02}_{-0.01}$ and $\langle f_{\rm esc}\rangle =0.05^{+0.02}_{-0.01}$ (Fig. \ref{fig:bayes_subsamples}). In either case, we find that any dependence of $f_{\rm esc}$ on $M_{\star}$, if one exists, is not a strong as the dependence on \wlya \ and $\beta$, suggesting that $M_{\star}$ is at best a secondary indicator of $f_{\rm esc}$.

\subsubsection{Intrinsic UV luminosity}
Finally, we divided the full galaxy sample at the median intrinsic (i.e. dust corrected) UV magnitude ($M_{\rm UV}=-21.8$), into UV-faint (median $M_{\rm UV}=-21.3$) and UV-bright (median $M_{\rm UV}=-22.4$) sub-samples, spanning intrinsic UV luminosities in the range $\mathbf{0.05\lesssim (L_{\rm{UV}}/L^{*}_{\rm{UV}})\lesssim2.5}$.
The maximum likelihood fitting technique returned 
values of $\langle f_{\rm esc}\rangle =0.18^{+0.06}_{-0.05}$ for the UV-faint galaxies and $\langle f_{\rm esc}\rangle=0.03^{+0.02}_{-0.01}$ for the UV-bright galaxies (Fig. \ref{fig:sampleChi2}). The constraints returned by the Bayesian inference approach are consistent, with $\langle f_{\rm esc}\rangle =0.10^{+0.04}_{-0.03}$ and $\langle f_{\rm esc}\rangle =0.04^{+0.01}_{-0.01}$, respectively (Fig. \ref{fig:bayes_subsamples}). We note that the decision to focus on the intrinsic UV magnitude rather than observed UV magnitude was taken because the latter is heavily dust-attenuation dependent, especially at the bright end, complicating the physical interpretation.

A summary of the $\langle f_{\rm esc}\rangle$ constraints for the full galaxy sample and the four sample splits considered here is presented in Table \ref{table:fesc_constrains}. The constraints derived from the two fitting approaches are consistent to at least the $2\sigma$ level, across all sample splits investigated. Taken together, these results form a consistent picture, in which LyC photons preferentially escape from the same UV-faint, dust-free galaxies that are also the primary sources of \Lya escape. Across the range of physical properties probed by our sample, the typical escape fraction appears to roughly encompass $f_{\rm esc}\simeq0-0.2$, with a full sample average of $\langle f_{\rm esc} \rangle \simeq 0.07$.
While our current sample is limited in terms of statistics and dynamic range, our analysis clearly demonstrates the ability of our adopted technique to recover $\langle f_{\rm esc}\rangle$ trends from broad-band photometry. 

\subsection{Individual $\bmath{\mathcal{R_{\rm obs}}}$ detections}\label{subsec:individual_detections}
\begin{figure}
        \centerline{\includegraphics[width=\columnwidth]{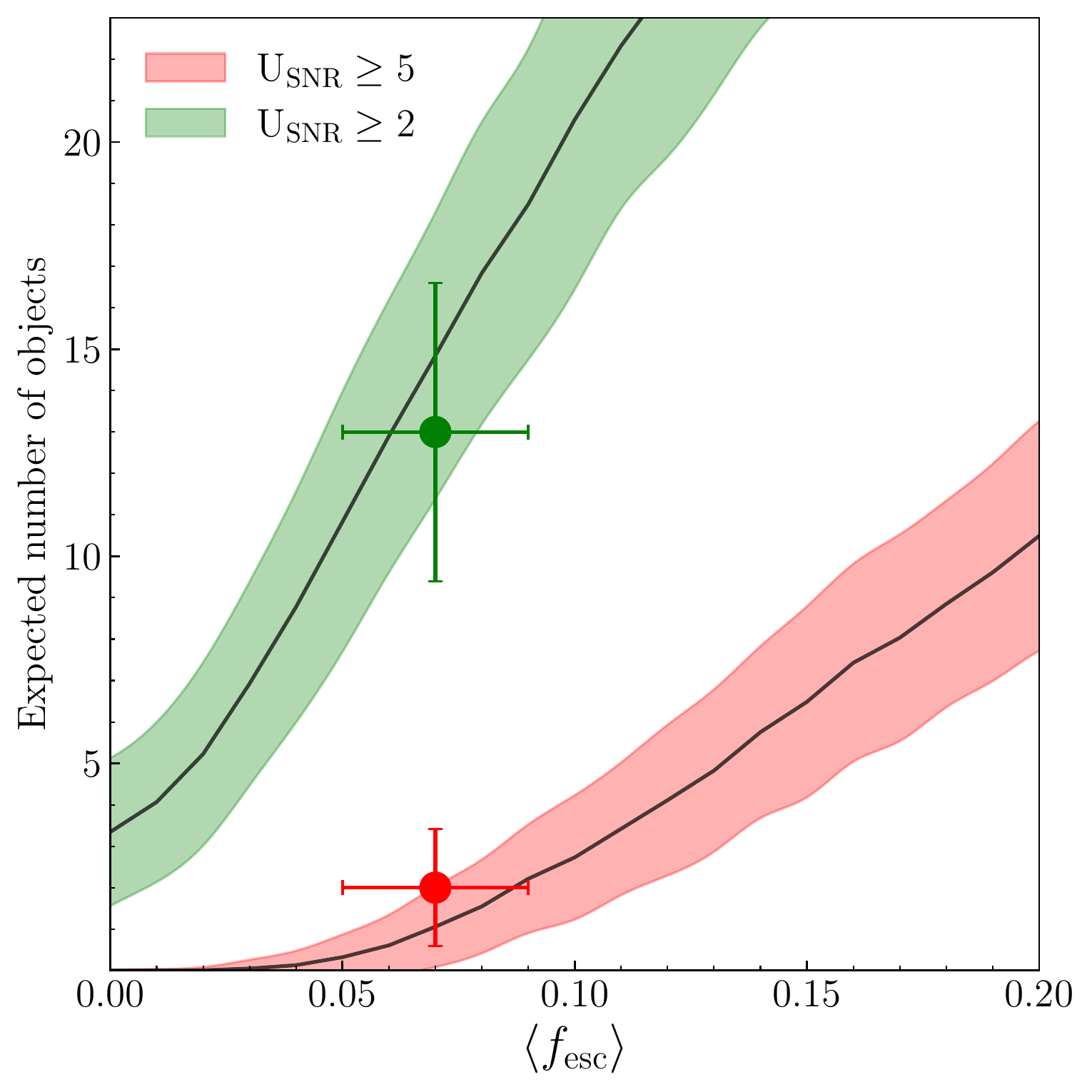}}
        \caption{The expected number of $U-$band flux detections in our sample as a function of $\langle f_{\rm esc} \rangle$ (see Section \ref{subsec:individual_detections}). The expected number of $U-$band flux detections at $\geq 2\sigma$ and $\geq 5\sigma$ based on our model are shown as the solid black lines, 
        with the $1\sigma$ uncertainties indicated by the shaded regions. The two data points indicate the observed number of $\geq 2\sigma$ and $\geq 5\sigma$ detections in our final sample.
        It can be seen that for our best-fitting value of 
        $\langle f_{\rm esc}\rangle \simeq 0.07$, the expected and observed number of $U-$band flux detections are in good agreement.}
        \label{fig:expectedDetections}
\end{figure}

The analysis presented here is primarily focused on constraining $\langle f_{\rm esc}\rangle$ for our galaxy sample based on modelling the shape of the $\mathcal{R}_{\rm obs}$ distribution. However, it is worth noting that two objects in our sample could be considered as robust LyC detections, having individual $U-$band flux measurements with $\geq 5\sigma$ significance. Both of these objects have been previously reported in the literature \citep{vanzella+10b,ji+20, saxena+21}.

In fact, the number of objects within our sample with a positive $U-$band flux detection provides a useful additional sanity check 
on the $\langle f_{\rm esc} \rangle$ value we derived for the full sample. 
We performed a simple test in which we constructed simulated samples as a function of average escape fraction, using the method described in Section~\ref{sec:model}. For each value of $\langle f_{\rm esc} \rangle$ in the range $0\leq~\langle~f_{\rm esc}~\rangle~\leq~0.2$ ($\Delta \langle f_{\rm esc} \rangle~=~0.01$) we produced 5000 simulated samples of 148 galaxies, and calculated the predicted number of $U-$band flux detections. As can be seen from Fig. \ref{fig:expectedDetections}, the number of $\geq 2\sigma$ and $\geq 5\sigma$ $U-$band flux detections we see in the real data is in good agreement with the model prediction for $\langle f_{\rm esc} \rangle=0.07\pm0.02$. 

\section{Discussion}\label{sec:discussion}

The results presented above clearly demonstrate that meaningful constraints on $\langle f_{\rm esc} \rangle$ at $z\simeq3-4$ can be obtained from broadband photometric measurements of the emergent LyC flux in the $U-$band.
In this section we begin by comparing our results to previous measurements in the literature at similar redshifts, before briefly considering the physical picture suggested by our results. We finish with a quantitative discussion of the various systematic uncertainties present in our study, suggesting avenues for future improvement.

\subsection{Literature comparison}
\begin{figure}
        \centerline{\includegraphics[width=\columnwidth]{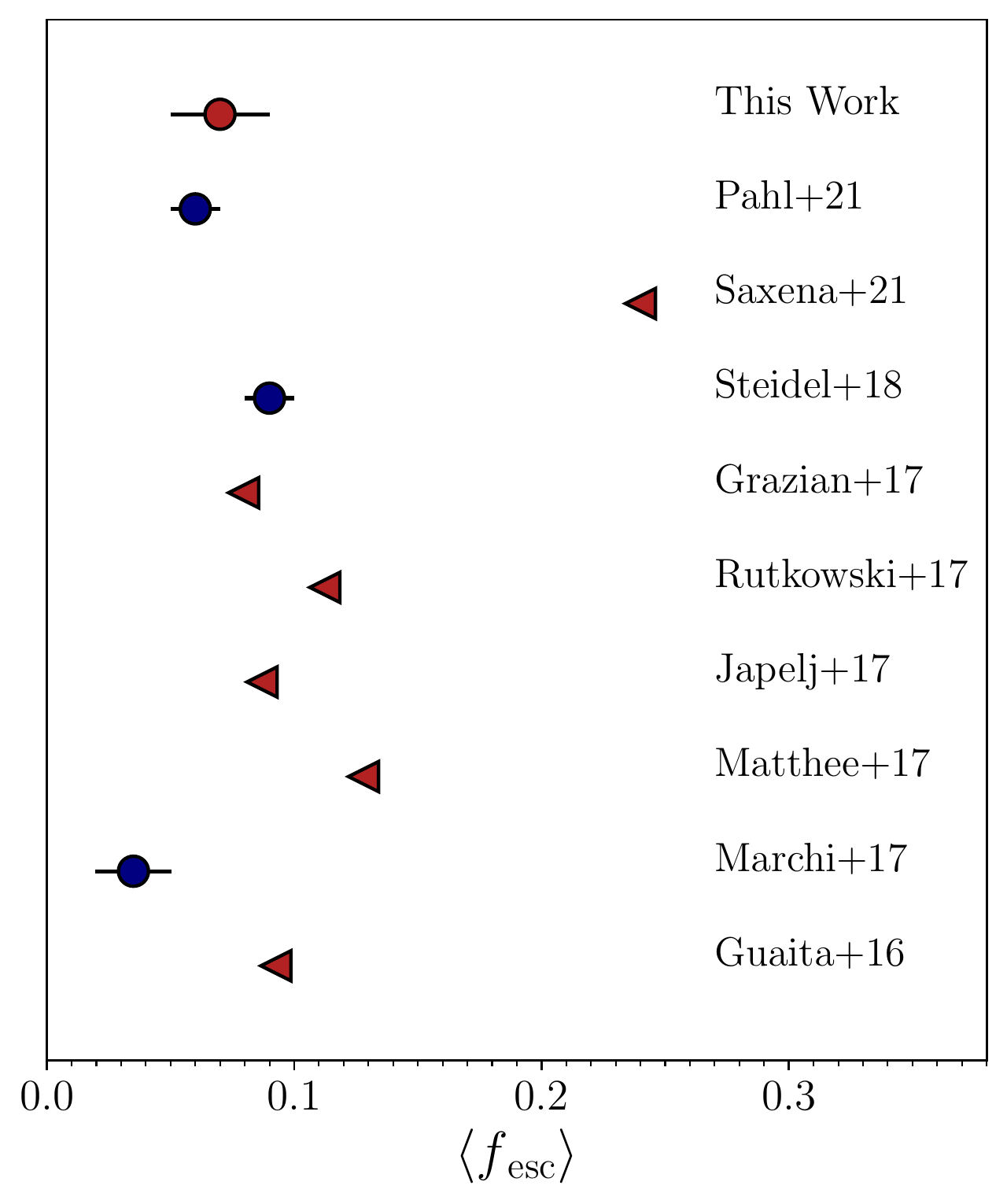}}
        \caption{A comparison between the results presented here and $\langle f_{\rm{esc}} \rangle$ measurements for $z\simeq3$ galaxy samples in the literature. Results based on photometry are shown in red and results based on spectroscopy are shown in blue. The circular markers represent studies that report a $\geq 2\sigma$ constraint on $\langle f_{\rm{esc}} \rangle$, while the triangular markers represent studies that report upper limits ($2\sigma$). Where necessary, we have converted relative escape fraction estimates to absolute escape fraction estimates assuming ${\rm{ E(B-V)}}=0.1$ and the \citet{calzetti+00} attenuation curve, as in \citet{mestric+21}. For \citet{grazian+17}, who derive constraints on $\langle f_{\rm{esc}} \rangle$ as a function of $M_{\rm{UV}}$, we plot the result for their $M_{\rm{UV}}\sim-19.7$ stack, closest to the median of our sample ($M_{\rm{UV}}\sim-20.2$).}
        \label{fig:litcomparison}
\end{figure}

In Fig.~\ref{fig:litcomparison} we show a comparison between the 
$\langle f_{\rm esc}\rangle$ constraints presented in this work and a selection of comparable studies of star-forming galaxies at $z\sim 2-4$ from the literature.
The literature compilation includes estimates of $\langle f_{\rm esc} \rangle$ derived from both spectroscopy and photometry. It can be seen that, prior to this work, the only statistical ($\geq 3\sigma$) measurements of $\langle f_{\rm esc} \rangle$ have come from deep spectroscopic analyses \citep[e.g.][]{marchi+17,steidel+18,pahl+21}.

The relative success of spectroscopic studies versus photometric studies indicated by Fig.~\ref{fig:litcomparison} is primarily due to the fact that spectroscopy enables a measurement of the LyC flux across a narrow bandpass, close to the intrinsic Lyman limit, where the optical depth to \hi \ is minimized \cite[e.g. the $880-910$ \AA \ window used by ][]{steidel+18}. For example, at $z=3.6$, the average IGM+CGM transmission integrated across the $U$-band filter is $\langle e^{-\tau_{\rm{U}}^{\rm H\textsc{i}}} \rangle=0.07$, compared to an average of $\langle e^{-\tau_{900}^{\rm H\textsc{i}}} \rangle=0.28$ across the $880-910$ \AA \ bandpass.

Despite this, the analysis presented here clearly demonstrates that it is possible to derive constraints from broad-band imaging that move beyond upper limits, and are comparable to those achieved from spectroscopy. 
However, to achieve this, large statistical samples, ultra-deep $U-$band imaging, accurate treatment of the IGM+CGM optical depth, and an analysis that exploits the full  $\mathcal{R}_{\rm obs}$ distribution, are all required.

It can be seen from Fig. \ref{fig:litcomparison} that our best-fitting value of $\langle f_{\rm esc}\rangle =0.07\pm0.02$ is fully consistent with the latest estimates from the VIMOS Ultra Deep Survey presented in \citet{marchi+17}, and the Keck Lyman Continuum Spectroscopic Survey presented in \citet{steidel+18} and \citet{pahl+21}.
Moreover, these results are in quantitative agreement with the majority of previous upper limits reported from broadband imaging studies, which typically find $\langle f_{\rm esc} \rangle \lesssim 0.1$ at the $2 \sigma$ level.

\subsection{A physical picture}

Our results point towards a surprisingly simple physical picture, in which the observed distribution of the ionizing to non-ionizing UV flux ratio for our galaxy sample can be modelled as population with a single value of $\langle f_{\rm esc}\rangle=0.07\pm0.02$.
However, we have also found evidence that $\langle f_{\rm esc}\rangle$ varies as a function of galaxy properties; increasing with \wlya\ and decreasing with UV dust attenuation and intrinsic UV luminosity. 
These trends are in good agreement with recent results from spectroscopic studies at similar redshifts \citep{steidel+18,pahl+21}, and have a number of important implications. Most importantly, our result suggest that low-dust, UV-faint galaxies at $z\geq 6$ are plausibly capable of displaying the $\langle f_{\rm esc} \rangle \geq 0.1$ required to drive reionization \citep[][]{robertson+13,finkelstein+19}.

As far as our sample is concerned, it is clear from Fig.~\ref{fig:distngrid} and Fig.~\ref{fig:expectedDetections} that the observed data is fully consistent with our simple model. Indeed, 
based on our sample alone, there is no indication that the more complex pictures that have been suggested in the literature, in which LyC continuum emission is switched either `on' or `off' due to anisotropic dust/ISM distributions leading to line-of-sight effects, and/or stochastic star-formation histories \citep[e.g.][]{fletcher+19}, are required to explain the data. 

However, although our current sample does not justify a more complex physical model statistically, it is clear that the true underlying $f_{\rm esc}$ distribution is likely to be significantly more complicated. In the future, larger sample sizes and deeper photometry should make it possible to fit more complex underlying $f_{\rm esc}$ distributions and quantitatively compare them to our simple model using Bayesian model selection techniques.

Interestingly, in contrast to the clear correlations between $f_{\rm esc}$ and \wlya, $\beta$ and $M_{\rm UV}$, our results do not show evidence for a strong trend between $f_{\rm esc}$ and galaxy stellar mass. The lack of a clear $f_{\rm esc}-M_{\star}$ trend is perhaps surprising given the known correlations between $M_{\star}$, \wlya \ and $\beta$ \citep[e.g.][]{mclure2018_irxbeta, cullen+20}; however, our results suggest while strong \lya{} emission, low dust attenuation, and faint intrinsic UV luminosity can be considered primary indicators of $f_{\rm esc}$, any $f_{\rm esc}$ trend with $M_{\star}$, if present, is secondary and `washed-out' by the scatter in the $M_{\star}-$\wlya, $M_{\star}-\beta$ and $M_{\star}-M_{\rm UV}$ relations.
This apparent lack of a clear $f_{\rm esc}-M_{\star}$ correlation is in good agreement with recent results for galaxies at $z\simeq0.4$ reported in \citet{izotov+21}.
However, it is important to note that of the four properties physical properties considered here, $M_{\star}$ is the most model-dependent, and therefore subject to the largest number of systematic uncertainties.

Finally, we note that our results are qualitatively in agreement with the results of \citet{reddy+16}, who found that the ionizing escape fraction is driven predominantly by changes in the covering fraction of $\rm{H}\textsc{i}$ ($f_{\rm cov}(\rm{H}\textsc{i})$), with lower $f_{\rm cov}(\rm{H}\textsc{i})$ corresponding to higher $f_{\rm esc}$ \citep[see also;][]{gazanges2020,reddy+21,saldana-lopez22}.
Crucially, \citet{reddy+16} showed that dust covering fraction increases with $f_{\rm cov}(\rm{H}\textsc{i})$, implying that $f_{\rm esc}$ will decrease towards galaxies with lower dust attenuation, consistent with our findings.
Indeed, a fundamental correlation between $f_{\rm esc}$ and $f_{\rm cov}(\rm{H}\textsc{i})$ (or dust) is well-motivated from a theoretical point of view, and naturally explains the resulting $f_{\rm esc}$ trends with \wlya \ and $\beta$.

\subsection{Systematic uncertainties}

Before concluding, it is worth considering the systematic impact of some of the key choices made in our analysis and understanding the effect that they have on our results.

One simplifying assumption we made was the choice of a single underlying SPS model for 
the full sample, which fixes the intrinsic ratio of ionizing to non-ionizing UV flux (Section \ref{subsec:intrinsic_sed}). Although this single model is physically motivated, being derived from a full-spectrum fit to the stacked VANDELS spectrum of our full galaxy sample, other choices for the intrinsic underlying SED would systematically alter the value of $\mathcal{R}_{\rm int}$, and hence $\langle f_{\rm esc} \rangle$. However, it is important to note that our choice of the BPASS binary models yields larger values of $\mathcal{R}_{\rm int}$ than most other available SPS models, and in that sense most alternative models would result in marginally larger $\langle f_{\rm esc} \rangle$ estimates. However, this is a relatively small systematic effect, for example, assuming the Starburst99 models \citep{leitherer-2010}, with the same star-formation history and IMF, increases $\langle f_{\rm esc} \rangle$ by a factor of $\simeq 1.1$.

We have also assumed constant star-formation histories on a 100 Myr timescale, a common assumption when modelling the stellar populations of $z\simeq2-3$ star-forming galaxies \citep[e.g.,][]{steidel+16, cullen+19}. Alternative choices for the star-formation history have only a minor effect on $\mathcal{R}_{\rm int}$ ($<10$ per cent), as long as the assumption of a constant star-formation rate is valid on timescales $\gtrsim 50$ Myr. Systematically younger ages would increase $\mathcal{R}_{\rm int}$ and hence decrease the resulting $\langle f_{\rm esc} \rangle$ (at the $\gtrsim 20$ per cent level). However, there is currently no strong observational evidence for star-formation timescales $< 50$ Myr at the typical stellar mass of our sample.

Another simplifying assumption we made was to adopt a single dust law, with a slope of $\delta=-0.25$, motivated by a comparison between the intrinsic model and the stacked spectrum of the full sample (Section \ref{sec:model}). A $\delta=-0.25$ slope is intermediate between the commonly-adopted Calzetti ($\delta=0$) and SMC curves ($\delta=-0.5$). Adopting a dust curve as steep as the SMC extinction curve would {\it increase} our derived $\langle f_{\rm esc} \rangle$ values by a factor of $\simeq1.5$.  In contrast, adopting an attenuation curve as grey as the \citet{calzetti+00} starburst law would {\it decrease} our derived $\langle f_{\rm esc} \rangle$ values by a factor of $\simeq1.7$.

In reality, each galaxy in our sample has a unique metallicity, star-formation history and dust attenuation curve, which could in principle be incorporated directly into the determination of $\langle f_{\rm esc} \rangle$ in a future analysis. Nevertheless, first-order estimates of the potential systematic effects suggest the range of $\langle f_{\rm esc} \rangle$ for the full sample would remain within the range $0.03 \leq \langle f_{\rm esc} \rangle \leq 0.1$. 

\section{Conclusions}\label{sec:conclusions}

We have presented the results of a study aimed at constraining the average Lyman-continuum escape fraction $\langle f_{\rm esc} \rangle$ of star-forming galaxies at $z\simeq3.5$.
After performing a careful selection against line-of-sight contamination and AGN interlopers, we assembled a sample of $148$ galaxies at $3.35 \leq z_{\rm spec} \leq 3.95$ from the VANDELS spectroscopic survey \citep{mclure+18, garilli+21} . 

Using a combination of ultra-deep, ground-based, $U-$band imaging and Hubble Space Telescope $V-$band imaging, we were able to robustly measure $\mathcal{R_{\rm obs}}$ $=(L_{\rm LyC}/L_{\rm UV})_{\rm obs}$ for each galaxy. By fitting the $\mathcal{R_{\rm obs}}$ distribution of our full sample, we were able to derive consistent constraints on $\langle f_{\rm esc} \rangle$, using two different fitting techniques. Both techniques were based upon the assumption that a single value of $f_{\rm esc}$ could be applied to the full sample and utilised accurate Monte Carlo simulations to trace the full distribution of $\rm{H}\textsc{i}$ optical depths through the intervening IGM and CGM.

Splitting the sample in two, we investigated the evidence for trends between $\langle f_{\rm esc}\rangle$ and a number of physical properties, namely \wlya, $\beta_{\rm obs}$, $M_{\star}$ and intrinsic $M_{\rm UV}$. The main results of this study can be summarised as follows:

\begin{enumerate}

    \item Fitting the $\mathcal{R}_{\rm obs}$ distribution for the full sample using a maximum-likelihood technique returns a best-fitting value of $\langle f_{\rm esc}\rangle~=~0.07\pm~0.02$ (Fig. \ref{fig:chi2_fullsample}). Models with $\langle f_{\rm esc}\rangle=0$ and $\langle f_{\rm esc}\rangle\geq 0.15$ are rejected at the $\geq 3\sigma$ level. 
    Using an independent Bayesian inference approach, we obtain a fully consistent value of $\langle f_{\rm esc}\rangle =0.05\pm 0.01$ (Fig. \ref{fig:bayes_result}).
    This result represents the first significant measurement ($\geq 3\sigma$) of $\langle f_{\rm esc}\rangle$ from ground-based broadband imaging at $z>3$. 

    \item Splitting the full sample into sub-samples based on various physical properties, we find evidence that $\langle f_{\rm esc}\rangle$ positively correlates with $W_{\lambda}(\rm{Ly}\alpha)$, but anti-correlates with intrinsic UV luminosity and UV dust attenuation. We find that the high $W_{\lambda}(\rm{Ly}\alpha)$, low intrinsic UV luminosity and low dust attenuation sub-samples all return best-fitting $\langle f_{\rm esc}\rangle$ values in the range $0.12~\leq~\langle f_{\rm esc} \rangle~\leq~0.22$ (see Fig.~\ref{fig:sampleChi2} and Fig.~\ref{fig:bayes_subsamples}).
    
    \item In contrast, splitting the sample by $M_{\star}$ yields a weak/non-existent trend between $M_{\star}$ and $f_{\rm esc}$.
    Our results suggest that $M_{\star}$ is, at best, a secondary indicator of $f_{\rm esc}$.
    Therefore, any trend between $f_{\rm esc}$ and $M_{\star}$ is likely the result of the known correlations between $M_{\star}$ and the other stronger indicators: \wlya, $\beta_{\rm obs}$ and $M_{\rm UV}$. 
    
    \item Overall, the results of the sub-sample splits suggest that the young, low metallicity, dust-free galaxies expected to be common at $z\geq 6$ are likely to display $\langle f_{\rm esc}\rangle\geq 0.1$, the threshold often quoted as necessary for them to drive cosmic reionization.
    
    \item The agreement between our simple model and the observed data (Fig. \ref{fig:distngrid} and Fig. \ref{fig:expectedDetections}) suggests that, at least for our sample, a more complicated model of the underlying $f_{\rm esc}$ distribution is not statistically justified. Nevertheless, it is clear that the true underlying $f_{\rm esc}$ distribution is likely to be significantly more complicated than our simple model. Therefore, it would clearly be desirable to expand the modelling performed here to larger galaxy samples with deeper photometry, in order 
    to explore more complicated $f_{\rm esc}$ distributions and to improve the significance of the correlations between $f_{\rm esc}$ and various galaxy properties.
    
\end{enumerate}

\section*{Acknowledgements}
The authors would like to acknowledge Alice Shapley and Charlotte Mason for useful discussions. MLl acknowledges support from the National Agency for Research and Development (ANID)/Scholarship Program/Doctorado Nacional/2019-21191036. ASL acknowledges support from the Swiss National Science Foundation.
This research made use of Astropy, a community-developed core Python package for Astronomy \citep{astropy13,astropy18},  NumPy \citep{numpy20} and SciPy \citep{scipy20}, Matplotlib \citep{matplotlib07}, IPython \citep{ipython07} and NASA’s Astrophysics Data System Bibliographic Services

\section*{Data Availability}
The VANDELS survey is a European Southern Observatory Public Spectroscopic Survey. The full spectroscopic dataset, together with the complementary photometric information and derived quantities are available from \url{http://vandels.inaf.it}, as well as from the ESO archive \url{https://www.eso.org/qi/}. The $U-$band imaging data is also publicly available from the ESO archive, with information found at \url{https://archive.eso.org/cms/eso-data/data-packages/goods-vimos-imaging-data-release-version-1-0.html}. The {\it HST} imaging is publicly available from the Hubble Legacy Fields data release at the STSCI archive from \url{https://archive.stsci.edu/prepds/hlf/#data-products}. For the purpose of open access, the author has applied a Creative Commons Attribution (CC BY) licence to any Author Accepted Manuscript version arising from this submission.



\bibliographystyle{mnras}
\bibliography{vandels_lyc} 




\appendix

\section{Details of the IGM+CGM opacity models}\label{app:igm_model}

To generate our IGM+CGM Monte Carlo models we adopted the parameterization for the column density and redshift distribution of \hi \ clouds outlined in \citet{steidel+18}.
In this prescription, the number of absorbers of column density between $N_{\mathrm{H I, max}}$ and $N_{\mathrm{H I, min}}$, and redshift between $z1$ and $z2$, is given by
\begin{equation}
N_{\mathrm{abs}}=\int_{N_{\mathrm{H I, min}}}^{N_{\mathrm{H I, max}}}\int_{z1}^{z2}N_{\mathrm{H I}}^{-\beta}A(1+z)^{\gamma}dN_{\mathrm{H I}}dz,
\end{equation}
where $\beta$ and $\gamma$ are power-law exponents describing the column destiny and redshift distribution of the number of absorbers and $A$ is a constant chosen to match the observations of \citet{rudie+13} for galaxies in the redshift range $2.0 \lesssim z \lesssim 2.8$.
The distributions are split into three regimes: low-density IGM, high-density IGM, and CGM, with the relevant parameters for each regime given in Table 11 of \citet{steidel+18}.

To generate sightlines at a given redshift, absorbers within the column density range $12<\mathrm{log}(N_{\mathrm{H I}})<21$ were drawn randomly from the appropriate distribution function.
For an absorber with column density $N_{\mathrm{H I}}$ and Doppler parameter $v_D$, the optical depth as a function wavelength is given by
\begin{equation}
\tau(\lambda) = N_{\mathrm{HI}} \bigg[ \sigma_{\mathrm{LyC}}(\lambda) + \sum{\sigma_i(\lambda)} \bigg],
\end{equation}
where $\sigma_{\mathrm{LyC}}(\lambda)$ is the absorption cross-section of the Lyman continuum and $\sigma_i(\lambda)$ is the absorption cross-section of the $i$th Lyman series line.
The cross section for the Lyman continuum is approximated to
\begin{equation}
\sigma_{\mathrm{LyC}}\simeq6.3 \times 10^{-18} \bigg( \frac{\lambda}{\lambda_{\mathrm{LyC}}} \bigg)^3,
\end{equation}
where $\lambda \leq \lambda_{\mathrm{LyC}}$ (i.e. when $\lambda > \lambda_{\mathrm{LyC}}$, $\sigma_{\mathrm{LyC}}=0$).

Following \citet{tepper-garcia+06}, the cross section for a given line in the Lyman series is given by
\begin{equation}
\sigma_i(\lambda) = \kappa_i H[a_i,x] \ \ \ [\mathrm{cm}^2],
\end{equation}
where
\begin{equation}
\kappa_i=\frac{\sqrt{\pi}e^2\lambda_i^2f_i}{m_e c^2 \Delta \lambda_D} \ \ \ [\mathrm{cm}^2],
\end{equation}
where $f_i$ is the oscillator strength of the strength of the transition, $\lambda_{i}$ is the wavelength of the transition in cm. Note that the units here are in the cgs system (e.g. $e=4.803 \times 10^{-10}$ Fr, $m_e=9.109\times10{-28}$ g; $c=2.998\times10^{10}$ cms$^{-1}$). $\Delta \lambda_D$ is the Doppler width of the line given by
\begin{equation}
\Delta \lambda_D = \frac{v_D}{c}\lambda_i.
\end{equation}
$H[a_i,x]$ is the Voigt-Hjerting profile which describes the shape of the line, the analytic approximation to this function given by \citet{tepper-garcia+06} is
\begin{align}
H[a_i,x]&=H_0-a_i/\sqrt{\pi}/x^2[H_0^2(4x^2x^2+7x^2 \nonumber\\
&+4+Q)-Q-1],
\end{align}
where $x=(\lambda-\lambda_i)/\Delta \lambda_D$, $H_0 \equiv e^{-x^2}$, and $Q \equiv 1.5/x^2$.
The dimensionless damping parameter $a_i$ is given by
\begin{equation}
a_i=\frac{\lambda_i^2 \Gamma_i}{4 \pi c \Delta \lambda_D},
\end{equation}
where $\Gamma_i$ is the damping constant, or the reciprocal of the mean lifetime of the transition.
The atomic data for the Lyman series transitions were taken from the NIST Atomic Spectra Database\footnote{https://physics.nist.gov/PhysRefData/ASD/lines$\_$form.html.} and we calculated up to the 40th transition.


\bsp	
\label{lastpage}
\end{document}